\documentclass[fleqn,usenatbib,useAMS]{mnras}
\usepackage{graphicx}   
\usepackage{amsmath}    
\usepackage{amssymb}    
\usepackage{multicol}        
\usepackage{bm}         
\usepackage{pdflscape}  
\usepackage[T1]{fontenc}
\usepackage{ae,aecompl}
\usepackage{newtxtext,newtxmath}

\newcommand{\logg} {\log \textsl{\textrm{g}}}
\newcommand{\qh} {q({\rm H})}

\newcommand{\Te} {T_{\rm eff}}

\newcommand{\msun} {$M_\odot$}

\newcommand\gta{\lower 0.5ex\hbox{$\buildrel > \over \sim\ $}} 
\newcommand\lta{\lower 0.5ex\hbox{$\buildrel < \over \sim\ $}} 

\newcommand{\ha} {$\rm{H}{\alpha}$}
\newcommand{\hb} {$\rm{H}{\beta}$}
\newcommand{\hgamma} {H$\gamma$}

\newcommand{\lsun} {$L_{\odot}$}

\newcommand{\logh} {\log {\rm H/He}}
\newcommand{\logca} {\log {\rm Ca/He}}

\title[Cool White Dwarfs in Gaia]{A Spectro-photometric Analysis of Cool White Dwarfs in the Gaia and Pan-STARRS Footprint}
\author[Caron, Bergeron, Blouin, and Leggett]
{Alexandre Caron$^1$, P. Bergeron$^1$, Simon Blouin$^2$, and S. K. Leggett$^3$\\
$^1$D\'epartement de Physique, Universit\'e de Montr\'eal, C.P. 6128, Succ. Centre-Ville, Montr\'eal, QC H3C 3J7, Canada\\
$^2$Department of Physics and Astronomy, University of Victoria, Victoria, BC V8W 2Y2, Canada\\
$^3$Gemini Observatory/NSF's NOIRLab, 670 N. A'ohoku Place, Hilo, HI 96720, USA\\
}

\begin{document}
\label{firstpage}
\pagerange{\pageref{firstpage}--\pageref{lastpage}}
\maketitle

\begin{abstract}

We present a spectro-photometric analysis of 2880 cool white dwarfs
within 100 pc of the Sun and cooler than $\Te\sim10,000$~K, with
$grizy$ Pan-STARRS photometry and Gaia trigonometric parallaxes
available. We also supplement our data sets with near-infrared {\it
  JHK} photometry, when available, which is shown to be essential for
interpreting the coolest white dwarfs in our sample. We perform a
detailed analysis of each individual object using state-of-the-art
model atmospheres appropriate for each spectral type including DA, DC,
DQ, DZ, He-rich DA, and the so-called IR-faint white dwarfs. We
discuss the temperature and mass distributions of each subsample, as
well as revisit the spectral evolution of cool white dwarfs. We find
little evidence in our sample for the transformation of a significant
fraction of DA stars into He-atmosphere white dwarfs through the
process of convective mixing between $\Te=10,000$~K and $\sim$6500 K,
although the situation changes drastically in the range
$\Te=6500$--$5500$~K where the fraction of He-atmosphere white dwarfs
reaches $\sim$45\%. However, we also provide strong evidence that at
even cooler temperatures ($\Te\lesssim5200$~K), most DC white dwarfs have H
atmospheres. We discuss a possible mechanism to account for this
sudden transformation from He- to H-atmosphere white dwarfs involving
the onset of crystallization and the occurrence of magnetism. Finally,
we also argue that DQ, DZ, and DC white dwarfs may form a more
homogeneous population than previously believed.

\end{abstract}


\section{Introduction}

The determination of the fundamental parameters of white dwarfs --
effective temperature, radius, mass, surface gravity, chemical
composition, and cooling age -- is of utmost importance for our
understanding of their global properties such as their temperature and
mass distributions, and their spectral evolution as well. The use of
white dwarfs as cosmochronometers, whether it is through the study of
their luminosity function or just by finding the oldest members of a
given population, also requires precise and accurate stellar
parameters. There are basically two main techniques that have been
applied to large sample of white dwarfs. The first one is the
spectroscopic technique
\citep{BSL92,liebert05,koester09,gianninas11,tremblay11}, where
optical spectra are compared to the predictions of detailed synthetic
spectra to measure both the effective temperature, $\Te$, and surface
gravity, $\logg$ (and in some instances the abundance of trace
elements), which are then converted into stellar mass using
evolutionary models. The second technique is the photometric technique
\citep{BRL97,hollands18, tremblay19b}, where magnitudes are
converted into average fluxes in several bandpasses, and compared to
the prediction of model atmospheres, synthetic photometry in this case
\citep{holberg06}. Here, $\Te$ and the solid angle $\pi(R/D)^2$ are
the fitted parameters, where $R$ is the stellar radius and $D$ is the
distance from Earth, which can be obtained from trigonometric parallax
measurements.

Over the years, we have observed a constant evolution of observational
data and theoretical models, with sometimes one aspect being ahead of
the other. For instance, \citet{BSL92} applied the spectroscopic
technique to a sample of only 129 DA stars using model spectra that
included the Hummer-Mihalas occupation probability formalism
\citep{hummer88}, which allowed for the first time realistic
quantitative measurements of the atmospheric parameters. All observed
spectra in those days were secured individually using single-slit
spectroscopy. In more recent years, the Sloan Digital Sky Survey (SDSS), with its very
efficient multi-fiber spectrograph, has revolutionized the field by
providing tens of thousands white dwarf spectra, and uncovered
several new spectral classes \citep{kleinman13,kepler19}. But detailed
analyses of such large white dwarf samples has also revealed
problems with the microphysics of model spectra (see, e.g.,
\citealt{genest19a}). 

Similarly, \citet{BRL97,BLR01} applied the photometric technique to
over 200 white dwarfs by measuring optical {\it BVRI} and
near-infrared {\it JHK} photometry of individual stars, with only 152
with trigonometric parallax measurements available. Their model
atmospheres relied on pure hydrogen, pure helium, or mixed H/He
compositions, which could lead to large uncertainties when applied to
DQ and DZ white dwarfs. \citet{dufour05,dufour07} analysed only years
later the same DQ and DZ stars using detailed model atmospheres
including traces of carbon and heavier elements. Now the amount of
photometric and astrometric data required to apply the photometric
technique has sky rocketed, thanks to large photometric surveys such
as the SDSS and Pan-STARRS (Panoramic Survey Telescope And Rapid
Response System; \citealt{chambers16}, \citealt{tonry12}), and most
importantly the Gaia astrometric mission, which provided precise
astrometric and photometric data for $\sim$260,000 high-confidence
white dwarf candidates \citep{gentile19,gentile21}. Such large data
sets have recently been equally matched by improvements with model
atmospheres, in particular at the cool end of the white dwarf sequence
where high atmospheric densities are reached
\citep{kowalski06,kowalski07,blouin17,blouin18a,blouin18b,blouindufour19}.
In parallel, similar efforts have been put forward to secure
spectroscopic observations of white dwarf candidates in the Gaia
survey \citep{kilic20,tremblay20,obrien22}.

Given these large amounts of data, it is common to attack the problem
by plowing through the photometric data sets and use the photometric
technique by making sound assumptions regarding the atmospheric
composition of each white dwarf in the sample (see, e.g.,
\citealt{bergeron19}, \citealt{lopez22}). However, there are some
limitations to this approach since ideally, one should perform a
tailored model atmosphere analysis of each individual object in the
sample, including also spectroscopic data. This is particularly
important for DQ and DZ stars since the presence of heavy elements in
the photospheric regions is known to affect the global atmospheric
structure and thus the derived atmospheric parameters \citep{dufour05,dufour07}. 

In this paper, we present such a detailed spectro-photometric analysis of
all spectroscopically identified white dwarfs within 100 pc from the
Sun, based on Gaia distances. We restrict our analysis to objects below
$\Te\sim10,000$~K and use a homogeneous set of model atmospheres to
study the DA, DC, DQ, DZ, He-rich DA, and the so-called IR-faint white
dwarfs present in our sample. Our secondary goal is to improve our
global effort to provide the best physical parameters of each white
dwarf in the Montreal White Dwarf
Database\footnote{http://montrealwhitedwarfdatabase.org/} (MWDD;
\citealt{dufour17}).

We present in Section \ref{sec:data} the selection of our white dwarf
sample and observational data sets, which are then analysed in Section
\ref{sec:analysis} using model atmospheres and fitting techniques
appropriate for each spectral type. We present selected results in
Section \ref{sec:results} including a discussion of the spectral
evolution of white dwarfs. Our conclusions follow in Section
\ref{sec:conclusion}.

\section{Observational Data}\label{sec:data}

\subsection{Sample Selection}\label{sec:sample}

We first selected all white dwarfs -- spectroscopically confirmed as
well as candidates -- found in the MWDD within 100 pc from the Sun. We
chose this distance limit to avoid problems related to interstellar
reddening. As specified in the MWDD, the white dwarf candidates in the
100 pc sample are selected from the Gaia Data Release 2
\citep[DR2,][]{gaia18}, allowing for the error on the parallax
measurement; we also included 27 objects for which the parallax
measurements became available only in the Gaia Early Data Release 3
(EDR3; \citealt{gaia22})\footnote{We estimated that the parallax
  zero-point corrections proposed by \citet{lindegren21} are of the
  order of 0.010 mas for our sample and were thus not applied.}.  The
recommendations described in \citet{lindegren18} were used to remove
non-Gaussian outliers in colours and absolute magnitudes, and the
sample was limited to objects with $>10\sigma$ significant parallax
(\citealt{bergeron19}, \citealt{kilic20}, MWDD). A cut in the $(G_{\rm
  BP}-G_{\rm RP}) - M_G$ plane was then performed to select the white
dwarf candidates (see also \citealt{kilic20}). Finally, we retained
only objects with Pan-STARRS {\it grizy} photometry measured in at
least 3 photometric bands, and restricted our analysis to white dwarfs
cooler than $\Te\sim10,000$~K based on the photometric analyses
described below.

Considering all the selection criteria, we end up with 8238 objects in
our sample, including 2880 spectroscopically confirmed white dwarfs
and 5358 white dwarf candidates.

\subsection{Photometric and Spectroscopic Data}\label{sec:photspec}

As mentioned above, we make use of the Pan-STARRS {\it grizy}
photometry available in the MWDD, although in the course of our
analysis, we discovered several erroneous, or simply missing,
photometric data sets in the MWDD, which have been corrected since.
We also supplement our {\it grizy} data sets with SDSS $u$ magnitudes,
which represent a powerful diagnostic for the study of cool white
dwarfs, as demonstrated in \citet{kilic20}. When the Pan-STARRS
photometry appeared in error, we rely instead on the SDSS {\it griz}
photometry. Finally, 7 white dwarfs in our sample are brighter than
$g=13.5$ (e.g., WD 0135-052 = EG 11 = L870-2 with $g=12.95$) and the
magnitudes are obviously saturated, in agreement with the limits
provided in \citet{magnier13}. For these objects we adopt instead the
optical {\it BVRI} photometry available in the MWDD.

\citet{BRL97,BLR01} discussed at length the importance of using
near-infrared {\it JHK} photometry for the analysis of cool white
dwarfs. \citet{blouin19b}, for instance, revisited the spectral
evolution of cool white dwarfs by restricting their sample only to
objects with infrared photometry available in the Two Micron Sky
Survey (2MASS). Here we go further by including all {\it JHK}
photometry published in the literature, and also matched the
spectroscopically confirmed white dwarf sample to the near-infrared
photometric catalogs provided by 2MASS, the Tenth Data Release of the
UKIRT Infrared Deep Sky Survey (UKIDSS), the UKIRT Hemisphere Survey
(UHS) JBand Data Release, and the Visible and Infrared Survey
Telescope for Astronomy (VISTA) public survey (see also
\citealt{leggett18}). We end up with about 92\% of the objects in our
sample having {\it JHK} photometry in at least one band (usually $J$),
61\% in at least two bands, and 51\% in all three.

Optical spectra are also used in our model atmosphere analysis to
constrain the atmospheric composition of the objects in our sample in
the case of DA and DC stars, or to measure the carbon and metal
abundances in the case of DQ and DZ stars, respectively. We rely here
on the spectra available in the MWDD, which have been secured from
various sources, but mostly from the SDSS.  For some objects without
spectra, we rely on the spectral types provided in the MWDD, which
come essentially from Simbad. Some of these spectral types can be
confirmed from the spectra published in the literature (see, e.g.,
\citealt{kawka07}), while others have published spectral types without
any actual spectra being displayed, and these should be considered more
uncertain. Also, in the course of our analysis, we revised some of the
published spectral types, and these are indicated in the results
presented below.

We summarize in Table 1 the observational data used in our analysis,
where we provide for each object the given J name based on Gaia
coordinates, Gaia ID (DR2; if not available the EDR3 is given
and marked by a star symbol in Table 1), spectral type, trigonometric
parallax measurement ($\varpi$), SDSS $u$ photometry, Pan-STARRS {\it
  grizy} photometry, and near-infrared {\it JHK} photometry. The
spectral types followed by an asterisk symbol correspond to newly
assigned types based on our analysis.

\begin{table*}
\centering
\caption{Observational Data. The spectral types followed by an
  asterisk symbol correspond to newly assigned types based on our
  analysis. This table, including uncertainties, is available in its
  entirety in machine-readable form as well as a pdf file.}
\begin{tabular}{lrcccccccccccc}
\hline
\hline
Name & Gaia ID (DR2/EDR3*) & Sp Type & $\varpi$ (mas) & $u$ & $g$ & $r$ & $i$ & $z$ & $y$ & $J$ & $H$ & $K$\\
\hline

J0000+0132 & 2738626591386423424 & DA & 14.95 & 16.65 & 16.23 & 16.34 & 16.48 & 16.64 & 16.75 & 16.23 & 16.14 & 16.24\\
J0000+1906 & 2774195552027050880 & DC & 9.800 & 23.29 & 20.32 & 19.73 & 19.52 & 19.43 & 19.34 & 18.65 & $\cdots$ & $\cdots$\\
J0001+3237 & 2874216647336589568 & DC & 10.15 & 20.42 & 19.49 & 19.16 & 19.06 & 19.04 & 19.01 & 18.32 & $\cdots$ & $\cdots$\\
J0001+3559 & 2877080497170502144 & DC* & 11.66 & 20.04 & 19.03 & 18.85 & 18.83 & 18.58 & 18.45 & 17.81 & $\cdots$ & $\cdots$\\
J0001$-$1111 & 2422442334689173376 & DC & 13.51 & 19.09 & 18.48 & 18.28 & 18.28 & 18.31 & 18.35 & 17.72 & $\cdots$ & 17.75\\
J0002+0733 & 2745919102257342976 & DA & 11.85 & 18.31 & 17.84 & 17.76 & 17.78 & 17.83 & 17.87 & 17.23 & 17.07 & 17.06\\
J0002+0733 & 2745919106553695616 & DAH & 12.19 & 18.52 & 18.05 & 17.98 & 18.01 & 18.09 & 18.15 & 17.59 & 17.44 & 17.51\\
J0002+1610 & 2772241822943618176 & DA & 9.817 & 19.55 & 18.93 & 18.71 & 18.64 & 18.67 & 18.72 & 17.98 & 17.66 & 17.54\\
J0002+6357 & 431635455820288128 & DC & 38.07 & $\cdots$ & 17.64 & 16.99 & 16.72 & 16.65 & 16.59 & 15.8 & 15.58 & 15.51\\
J0003+6512 & 432177373309335424 & DC & 10.66 & $\cdots$ & 17.59 & 17.61 & 17.73 & 17.84 & 17.94 & $\cdots$ & $\cdots$ & $\cdots$\\

\hline
\end{tabular}
\end{table*}

\subsection{Colour-Magnitude Diagrams}\label{sec:CMD}

\begin{figure}
\centering
\includegraphics[width=3.3in, clip=true, trim=0.in 0.in 0.0in 0.0in]{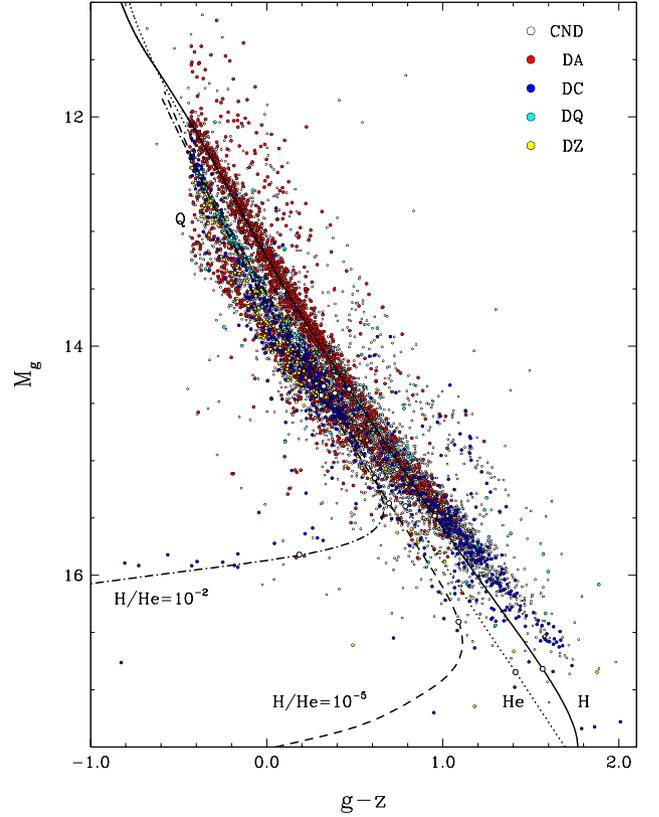}
\caption{Pan-STARRS $M_g$ versus $(g-z)$ colour-magnitude diagram for
  the 100 pc sample drawn from the MWDD. The various spectral types as
  well as the white dwarf candidates (CND) are identified with
  different colour symbols indicated in the legend. The various curves
  correspond to cooling sequences for 0.6 \msun\ CO-core models with
  pure H, pure He, $\logh=-5$ and $-2$ atmospheric compositions; white
  circles on each curve indicate $\Te=6000$ K, 5000 K, and 4000 K. The
  crystallization sequence (the so-called Q-branch), composed of a
  pile up of massive DA stars, is also indicated by the letter Q.
\label{color_mag_pan}}
\end{figure}

Before proceeding with a detailed model atmosphere analysis of the
observational material described above, it is worth examining a few
colour-magnitude diagrams. Figure \ref{color_mag_pan} shows the $M_g$
versus $(g-z)$ colour-magnitude diagram built from Pan-STARRS
photometry.  White dwarfs of various spectral types are identified by
different colours.  Also superposed are cooling sequences for 0.6
\msun\ CO-core models with pure H and pure He compositions, as well as
mixed compositions of $\log {\rm H/He}=-5$ and $-2$. Here and in the
remainder of this paper, we rely on the evolutionary models described
in \citet{bedard20} with C/O cores, $q({\rm He})\equiv \log M_{\rm
  He}/M_{\star}=10^{-2}$ and $q({\rm H})=10^{-4}$, which are
representative of H-atmosphere white dwarfs, and $q({\rm He})=10^{-2}$
and $q({\rm H})=10^{-10}$, which are representative of He-atmosphere
white dwarfs. As discussed at length by \citet{bergeron19}, a trace of
hydrogen is required to match the observed sequence of He-rich objects
in Figure \ref{color_mag_pan} with 0.6 \msun\ models, in particular
the DC and DZ white dwarfs. Such a trace of hydrogen is predicted by
the transformation of DA stars into He-rich white dwarfs through
convective mixing (see Figure 4 of \citealt{bedard22}). Note in
particular how the DZ stars overlap nicely with the DC white dwarfs in
this colour-magnitude diagram.

\begin{figure}
\centering
\includegraphics[width=3.3in, clip=true, trim=0.in 0.in 0.0in 0.0in]{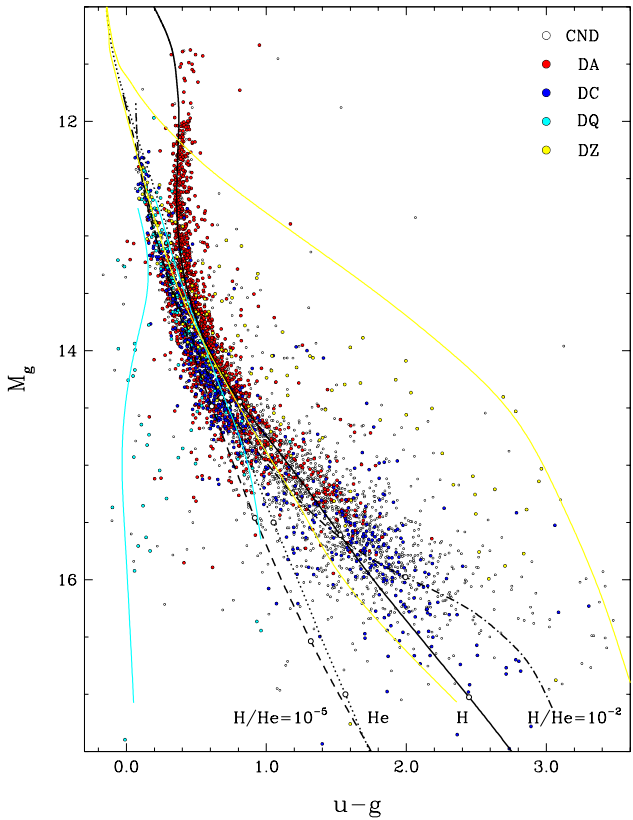}
\caption{Same as Figure \ref{color_mag_pan} but for the SDSS $M_g$
  versus $(u-g)$ colour-magnitude diagram.  Also shown are cooling
  sequences for 0.6 \msun\ CO-core models for DQ stars
  (cyan) with $\log {\rm C/He}=-5$ (left) and $-7.5$ (right), and for
  DZ stars (yellow) with $\log {\rm Ca/He}=-11.5$ (left) and $-7$
  (right). 
\label{color_mag_sdss}}
\end{figure}

The pure H sequence in Figure \ref{color_mag_pan} follows perfectly
the observed DA sequence, except at the faint end where the predicted
magnitudes fall below the observed sequence. The problem is
exacerbated at even lower luminosities where the observed DC sequence
-- presumably the extension of the DA sequence at lower temperatures
when \ha\ disappears -- lies significantly above the 0.6 \msun\ pure H
sequence. The problem is even worse if one assumes a pure He or mixed
H/He compositions, as seen in Figure \ref{color_mag_pan}. This
discrepancy between observed and predicted colours translates into low
inferred masses at low temperatures when the photometric method is
applied to this data set, as illustrated in Figure 12 of
\citet{bergeron19}, and even more so when assuming He-rich
atmospheres. The origin of this problem is still unknown.

Also observed at higher luminosities in Figure \ref{color_mag_pan} is
the crystallization sequence (the so-called Q-branch) composed of a
pile up of massive DA stars caused by the release of latent heat and
chemical fractionation, which decrease the cooling rate of a white
dwarf \citep{tremblay19}. In this particular color-magnitude diagram,
this sequence of massive DA stars is well separated from the normal
$\sim$0.6 \msun\ DA white dwarfs.

Also of interest in Figure \ref{color_mag_pan} is the small number of
white dwarfs at faint luminosities in the region where He-rich objects
should be located according to the model predictions. It is precisely
this dearth of cool DC stars at low luminosities with colours
consistent with pure helium compositions that led \citet{kowalski06}
to conclude that most cool, DC white dwarfs must have pure H
atmospheres. We come back to this point below. Perhaps even more
important is the existence of two distinct DC sequences, one at very
low luminosities that appears as the extension of the DA sequence, but
also a more luminous DC sequence, which terminates abruptly at
$\Te\sim5200$~K according to the theoretical models displayed in
Figure \ref{color_mag_pan}. Moreover, there are very few DC stars 
connecting these two DC sequences. We can also see a deficiency of
spectroscopically confirmed white dwarfs near $M_g\sim15$ and
$(g-z)\sim0.7$, but this appears to be an observational bias since
there are plenty of white dwarf candidates (CND) in the same region.

Figure \ref{color_mag_sdss} shows the $M_g$ versus $(u-g)$
colour-magnitude diagram built from the subsample of objects with SDSS
photometry (SDSS $g$ here). In this particular diagram, the DA and DC
sequences are well-separated at high luminosities as a result of the
Balmer jump present in H atmospheres. The 0.6 \msun\ pure H, pure He,
and mixed H/He sequences are also reproduced in this figure, but in
addition we show 0.6 \msun\ white dwarf models for DQ stars with $\log
{\rm C/He}=-7.5$ and $5.0$, and for DZ stars with $\log {\rm
  Ca/He}=-11.5$ and $-7.0$, which correspond roughly to values that
bracket the abundances measured for each spectral type in this
temperature range (see Figures 8 and 12 of
\citealt{coutu19}). Interestingly, the effect on the predicted colours
goes in the opposite direction when traces of carbon in DC stars and
metals in DZ stars are gradually increased, in agreement with the
location of these white dwarfs in this particular colour-magnitude
diagram. Note again here the lack of He-atmosphere white dwarfs at
faint luminosities, as well as the inability of pure H models to match
the observed faint DC sequence.

\section{Model Atmosphere Analysis}\label{sec:analysis}

\subsection{The Photometric Technique}\label{sec:phottech}

The physical parameters of the white dwarfs in our sample can be
measured using the so-called photometric technique, described at
length in \citet{BRL97}. Briefly, the optical and near-infrared
photometry are first converted into average fluxes in each bandpass
using the equations given in \citet{holberg06} appropriate for
magnitudes on the AB system for the SDSS $u$ and Pan-STARRS {\it
  grizy} optical photometry, or on the Vega system for the
near-infrared {\it JHK} photometry, with zero points calculated for
each filter system (2MASS, MKO, VISTA, Bessell, etc.). The synthetic
photometry is calculated for each model grid described below, by
integrating the monochromatic Eddington fluxes -- which depend on
$\Te$, $\logg$, and chemical composition -- over each filter
bandpass. The observed and model fluxes are related to each other by
the solid angle, $\pi(R/D)^2$, where $R$ is the stellar radius and $D$
is the distance from Earth, which can be obtained from the Gaia
trigonometric parallax measurements; given that our sample is limited
to objects with $>10\sigma$ significant parallax, it is justified to
simply assume here $D=\varpi^{-1}$. A $\chi^2$ value is calculated as
the difference between observed and predicted fluxes summed over all
bandpasses, and minimized using the nonlinear least-squares method of
Levenberg-Marquardt \citep{press86}, which is based on a steepest
decent method. Only $\Te$ and the solid angle $\pi (R/D)^2$ are
considered free parameters for an assumed chemical abundance, while
the uncertainties of both parameters are obtained directly from the
covariance matrix of the fit. The stellar radius is then converted
into mass using the evolutionary models described in Section
\ref{sec:CMD}. We use thick ($q({\rm H})=10^{-4}$) and thin ($q({\rm
  H})=10^{-10}$) H-layer models for H- and He-dominated atmospheres,
respectively. Bad photometric data points are excluded from our fits
after visual inspection, and displayed in red in the fits presented
below.

\subsection{DA White Dwarfs}\label{sec:DA}

\begin{figure}
\centering
\includegraphics[width=3.3in, clip=true, trim=0.5in 3.in 0.8in 0.0in]{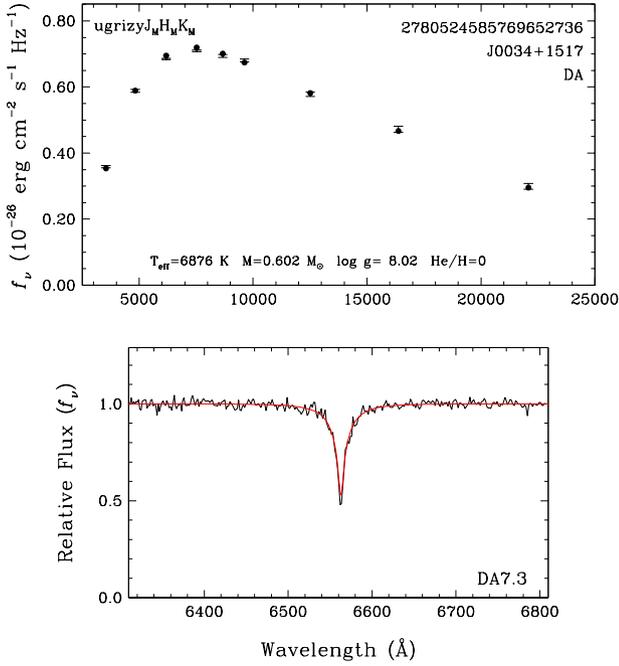}
\caption{Best fit to a typical DA star in our sample. The observed
  photometry is shown as error bars (1$\sigma$ uncertainty in all
  figures) and the photometric set used is indicated at the top left
  of the upper panel (see text). Our model atmosphere fit is shown as
  solid dots, with the parameters given in the upper panel. The lower
  panel shows the normalized optical spectrum (black), compared with
  the model spectrum (red) interpolated at the parameters obtained
  from the photometric solution.
\label{fit_DA}}
\end{figure}

Our model atmospheres for DA white dwarfs are described at length in
\citet{tremblay11} and references therein, and in \citet{blouin18a}
for $\Te<5000$~K, where additional nonideal effects are taken into
account. We assume a pure hydrogen atmospheric composition for the
1764 DA stars in our sample \footnote{We exclude the 17 He-rich DAs
  and the 4 DA + DQ unresolved systems, which are analysed further
  below.}. This is a perfectly legitimate assumption, as discussed in
\citet[][see also \citealt{blouin19b}]{BRL97}.  A fit to a typical DA
star, J0034+1517, is displayed in Figure \ref{fit_DA}. We also provide
the Gaia ID as well as the spectral type. The observed photometry is
shown as error bars (1$\sigma$ uncertainty) while the particular set
of photometry used is indicated at the top left of the upper panel. In
the case of the near-infrared {\it JHK} photometry, we use a lower
subscript to specify the particular system used (M: MKO, 2: 2MASS, B:
Bessell, V: VISTA).  Our best-fit model is represented by solid dots,
with the corresponding parameters given in the upper panel. The bottom
panel shows the optical spectrum, normalized to a continuum set to
unity, compared to the model spectrum (in red) interpolated at the
parameters obtained from the photometric solution and convolved with
the appropriate spectral resolution. Hence, it is not a formal fit,
but it is used instead to validate our photometric solution. An
observed discrepancy might be indicative of an unresolved double
degenerate binary (see Section \ref{sec:DD}), or of an incorrect
assumption about the chemical composition (see Section
\ref{sec:HeDA}). Depending on the spectral coverage, we show the
region near \ha, \hb, or even \hgamma, in cases where the red part of
the spectrum is contaminated by the presence of an M--dwarf
companion. We also provide in the lower panel the spectral type with
the temperature index defined as $50,400/\Te$.

Among our sample of DA white dwarfs, we also have 50 DAZ stars, where
in the present context we refer to white dwarfs with H-dominated
atmospheres and metal lines visible in their optical spectra, in
particular the Ca~\textsc{ii} H \& K doublet. If metal lines are
detected, which is the case for 34 DAZ stars in our sample, we
determine the Ca abundance by fitting the Ca~\textsc{ii} H \& K
doublet using a grid of model atmospheres with
$\Te=4500\ (500)\ 9000$~K and $\log {\rm Ca/H}=-10.5\ (0.5)\ $$-5.5$,
where the numbers in parentheses indicate the step size. We assume
chondritic abundance ratios with respect to Ca for other metals.
The same model grid is also used in Section \ref{sec:DZ} to
analyse cool H-atmosphere DZ and DZA white dwarfs.

The fits to all white dwarfs in our sample are provided as
Supplementary Material. For the fits to DA and DAZ white dwarfs, we
display both the pure H and mixed H/He solutions (with H as a trace
element, see Section \ref{sec:DC}); the entire spectrum is also shown
in a third panel at the bottom. In some cases there is no optical
spectrum available to us, and we simply rely on the published spectral
type.  In general, the predicted model spectra are in excellent
agreement with the observed absorption profiles. Exceptions include of
course the magnetic DA white dwarfs (see, e.g., J0505$-$1722) since
all our models are for non-magnetic stars (see also Section
\ref{sec:mag}). We note that \ha\ can be detected in DA stars as cool
as $\Te\sim4800$~K (see J0717+1125), although this depends on the
surface gravity of the white dwarf, and most importantly, on the
signal-to-noise ratio of the optical spectrum.

\subsection{He-rich DA White Dwarfs}\label{sec:HeDA}

Several DA stars in our sample have He-rich atmospheres.  The origin
of these so-called He-rich DA white dwarfs can be explained in terms
of DA stars with thin hydrogen layers that turn into a He-rich white
dwarf below $\Te\sim12,000$~K as a result of convective mixing. If the
object is hot enough, \ha\ might still be detectable as a shallow
absorption feature, heavily broadened by van der Waals interactions.
Classical He-rich DA white dwarfs include Ross 640, L745-46A, and GD
95, all three of which also show traces of metals (see Figure 14 of
\citealt{giammichele12}).  Many additional He-rich DA white dwarfs
have also been identified and analysed by \citet{rolland18}. As these
stars cool off, the \ha\ absorption feature rapidly falls below the
limit of visibility (see Figure 14 of \citealt{rolland18}).

\begin{figure}
\centering
\includegraphics[width=3.3in, clip=true, trim=0.5in 3.in 0.8in 0.0in]{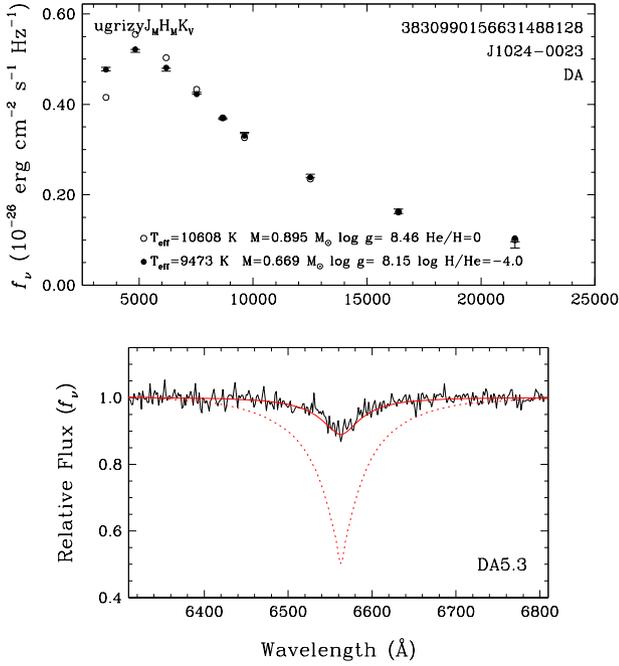}
\caption{Best fit to a typical He-rich DA star in our sample. The
  observed photometry is shown as error bars and the photometric set
  used is indicated at the top left of the upper panel. Our model
  atmosphere fits for pure hydrogen and mixed H/He compositions are
  shown with different symbols, with parameters given in the upper
  panel. The lower panel shows the normalized optical spectrum (black),
  compared with the model spectra (red) interpolated at the
  parameters obtained from the photometric solution (solid line: mixed
  H/He solution, dotted line: pure H solution). Note that the y-axis
  begins at 0.4.
\label{fit_HeDA}}
\end{figure}

The 17 He-rich DA white dwarfs in our sample have been fitted using mixed
H/He model atmospheres similar to those described above for DA stars,
but with $\logh = -5.0$, $-4.0$ (0.5) $-1.0$ (1.0) $+2.0$.  We use the
same photometric method as for DA stars, but this time we explore all
values of H/He individually and simply adopt the solution that most closely
matches the observed \ha\ profile.  A fit to a typical He-rich DA star
in our sample, J1024$-$0023, is displayed in Figure
\ref{fit_HeDA}. Here we show both the pure H and mixed H/He solutions,
as well as the corresponding predicted spectra at \ha. As can be seen,
the observations at \ha\ clearly rule out the pure H solution for this
object, whereas the mixed H/He solution constrained by \ha\ provides a
much better fit to the observed photometry, in particular in the $ugr$
passbands.

We also found in our sample what appears to be normal DA stars -- with
the complete Balmer series detected -- for which we achieve a much
better fit in the $ugr$ passbands by assuming a mixed composition of
$\logh = -1.0$ (see J1611+1322 and J2138+2309 in the Supplementary
Material), although we do not necessarily achieve better fits at \ha.
Also, two He-rich DA stars in our sample (J0103$-$0522, J1159+0007)
show asymmetric Balmer line profiles. \citet{tremblay20} suggested
that J0103$-$0522 was magnetic. However, the fact that J0103$-$0522
and J1159+0007 are almost identical twins, and that they both have
He-rich atmospheres, suggests instead that the asymmetric profiles are
the result of some unusual line broadening due to helium, unaccounted
for in our models, a hypothesis supported by the recent calculations
of \citet[][see their Figures 7 and 8]{spiegelman22}. Similar
asymmetric profiles can also be observed in the spectrum of GD 16 (see
Figure 18 of \citealt{limoges15}), another He-rich DA star (a DAZB
white dwarf in this case). Clearly, these peculiar white dwarfs
deserve further investigation.

\subsection{DQ White Dwarfs}\label{sec:DQ}

\begin{figure}
\centering
\includegraphics[width=3.3in, clip=true, trim=0.5in 3.in 0.8in 0.0in]{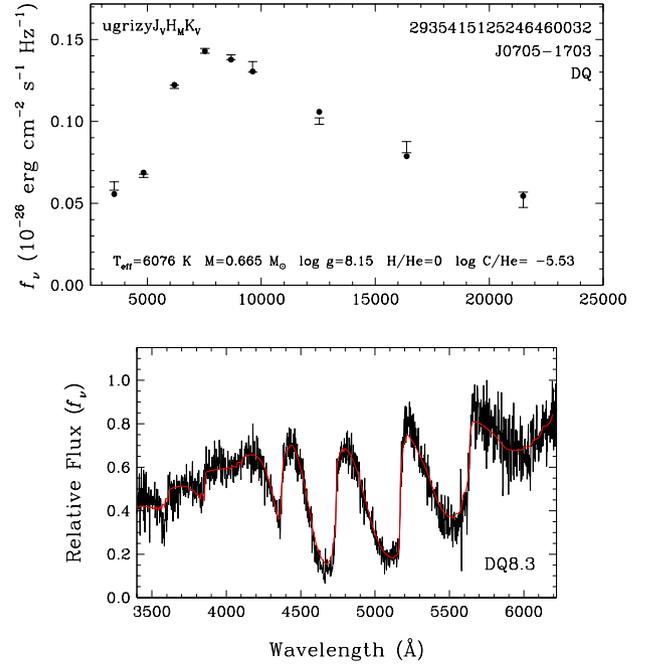}
\caption{Best fit to a typical DQ white dwarf in our sample. Upper
  panel: The observed photometry, indicated at the top left, is shown
  by error bars, while the model atmosphere fit is shown by solid
  dots, with parameters given at the bottom.  Lower panel: Optical
  spectrum (black) used to constrain the carbon abundance of the photometric
  solution, compared with the model spectrum (red).
\label{fit_DQ}}
\end{figure}

We fit the 146 DQ white dwarfs in our sample with spectra available
(including the four DA + DQ systems) using a hybrid spectroscopic and
photometric method outlined in \citet{dufour05}, although we rely here
on the improved models of \citet{blouin18a,blouin19b}, and
\citet{blouindufour19}. We first assume a carbon abundance and fit the
photometry using the same technique as before. We then fit the optical
spectrum -- neutral carbon lines or C$_2$ Swan bands -- to measure the
carbon abundance using the $\Te$ and $\logg$ values obtained from the
photometric solution.  We iterate this procedure until we reach an
internal consistency between the photometric and spectroscopic
solutions. A fit to a typical DQ star in our sample, J0705$-$1703, is
displayed in Figure \ref{fit_DQ}.

Some DQ stars in our sample also show a weak and broad \ha\ absorption feature
(J0916+1011, J0950+3238, and J1147+0747) and are thus genuine DQA
white dwarfs. In some cases, however, the presence of hydrogen lines is better
explained as the result of an unresolved DA+DQ double degenerate
binary (J0613+2050, J0928+2638, J1406+3401, and J1532+1356). In these cases,
not only the object appears overluminous, but hydrogen lines are also much sharper,
in contrast to the genuine DQA stars where \ha\ is heavily broadened by 
van der Waals interactions in a He-dominated environment. For the
three genuine DQA white dwarfs, we calculated a smaller set of model
atmospheres that also include hydrogen with $\logh = -5$, $-4$,
and $-3$, and constrain the hydrogen abundance to the closest value
set by the observations at \ha. For the four DA+DQ systems, we did not
attempt to deconvolve the individual components, and simply fit these
systems assuming a single DQ white dwarf; these are further discussed
in Section \ref{sec:DD}.

Given the temperature range explored in our analysis, most DQ stars in
our sample show only C$_2$ Swan bands, although some of the hottest
ones (e.g., J0859$+$3257) also show neutral carbon lines. In general,
the strong shift of the Swan bands, as modeled by
\citet{blouindufour19} and references therein, is remarkably well
reproduced (e.g., J1333+0016), although there are several exceptions
and problematic cases (e.g., J1113+0146, J1618+0611), already
discussed by Blouin \& Dufour (see their Section 3.1). Possible
explanations include problems with the simple model used to account
for the distortion of the Swan bands, or the presence of a strong
magnetic field, which has been confirmed by spectropolarimetric
measurements in some objects.

Finally, we could not help noticing that some of the DQ white dwarfs
in our sample appear to show some IR flux deficiency, the best example
of which is J1442+4013. Other examples in our sample include
J0033+1451, J0041$-$2221 (the well-studied LHS 1126;
\citealt{wickramasinghe82,bergeron94,kilic06b,blouin19b}), J0508$-$1450,
J1045$-$1906, J1247+0646, and J1614+1728. Indeed, our poor photometric
fits to these stars are reminiscent of the poor fits we obtain for the
so-called IR-faint white dwarfs (see Section \ref{sec:IRF}), when
analysed with pure H or pure He models instead of mixed H/He
models. In these cases, the IR-flux deficiency is the result of
collision-induced absorptions (CIA) by molecular hydrogen due to
collisions with helium. Perhaps the IR flux deficiency in these
peculiar DQ stars is also the result of absorption by molecular
hydrogen, thus offering the possibility of measuring H abundances in
cool DQ white dwarfs. Unfortunately, our attempts to fit these DQ
white dwarfs with mixed H/He/C model atmospheres were unsuccessful,
the main problem being that a strong CH absorption feature is always
predicted but not observed (with the possible exception of
J1442+4013). Clearly, these objects deserve further investigation.

\subsection{DZ White Dwarfs}\label{sec:DZ}

\begin{figure}
\centering
\includegraphics[width=3.3in, clip=true, trim=0.0in 2.5in 0.0in 0.0in]{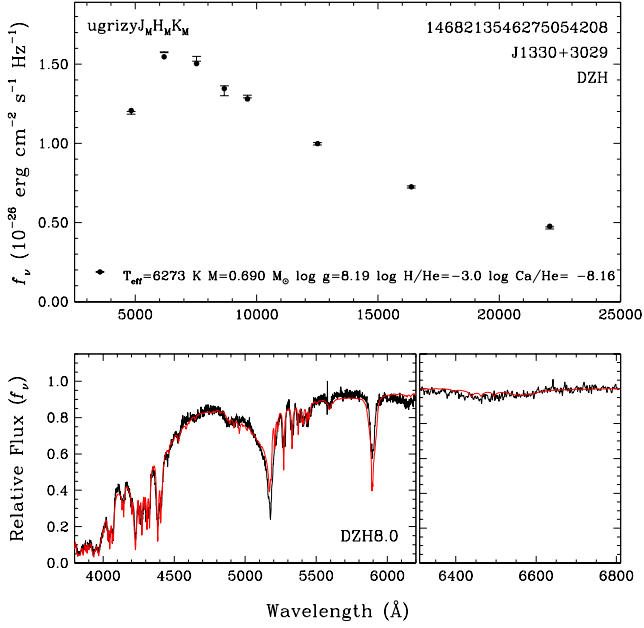}
\caption{Best fit to a typical DZ white dwarf in our sample. Upper
  panel: The observed photometry, indicated at the top left, is shown
  by error bars, while the model atmosphere fit is shown by solid
  dots, with parameters given at the bottom.  Lower panel: Optical
  spectrum (black) used to constrain the metal abundance (left) and hydrogen
  abundance (right) of the photometric solution, compared with the
  model spectra (red).
\label{fit_DZ}}
\end{figure}

The procedure used to fit the DZ and DZA white dwarfs in our
sample is similar to the approach used for DQ stars, with model
atmospheres described in \citet[][see also
  \citealt{coutu19}]{blouin18a}. The only exception here is that we
also rely on optical spectra around \ha, if available, to measure
(or constrain) the hydrogen abundance. Our solutions are provided in
terms of the Ca abundance ($\logca$), and we assume chondritic
abundance ratios with respect to Ca for other metals,
spectroscopically visible or not. Our model grid covers Ca abundances
of $\log {\rm Ca/He}=-12.0\ (0.5)\ $$-7.0$. As was the case for the
He-rich DA white dwarfs, we do not attempt to fit the hydrogen
abundance, but we simply explore the solutions with various hydrogen
abundances ($\logh=-5, -4, -3, -2, -1.5$, and $-1$) and adopt the value,
or limit, consistent with the \ha\ spectrum. A fit to a typical DZ
star in our sample, J1330+3029, is displayed in Figure
\ref{fit_DZ}. Even though the object shown here shows various metallic
absorption features, most DZ stars in our sample only show the
Ca~\textsc{ii} H \& K doublet.

We have 9 DZ stars in our sample without optical spectra available to
us.  For 4 of those, we rely on Ca abundance published in the
literature (J0124$-$2229, J0512$-$0505, J0823+0546, and J2343$-$1659),
while we fit the remaining 5 (J0153+0911, J0733+2315, J1442+0635,
J1551+1439, and J2147$-$2910) using the procedure for DC stars
described in Section \ref{sec:DC}.

\begin{figure}
\centering
\includegraphics[width=3.3in, clip=true, trim=0.0in 2.5in 0.0in 0.0in]{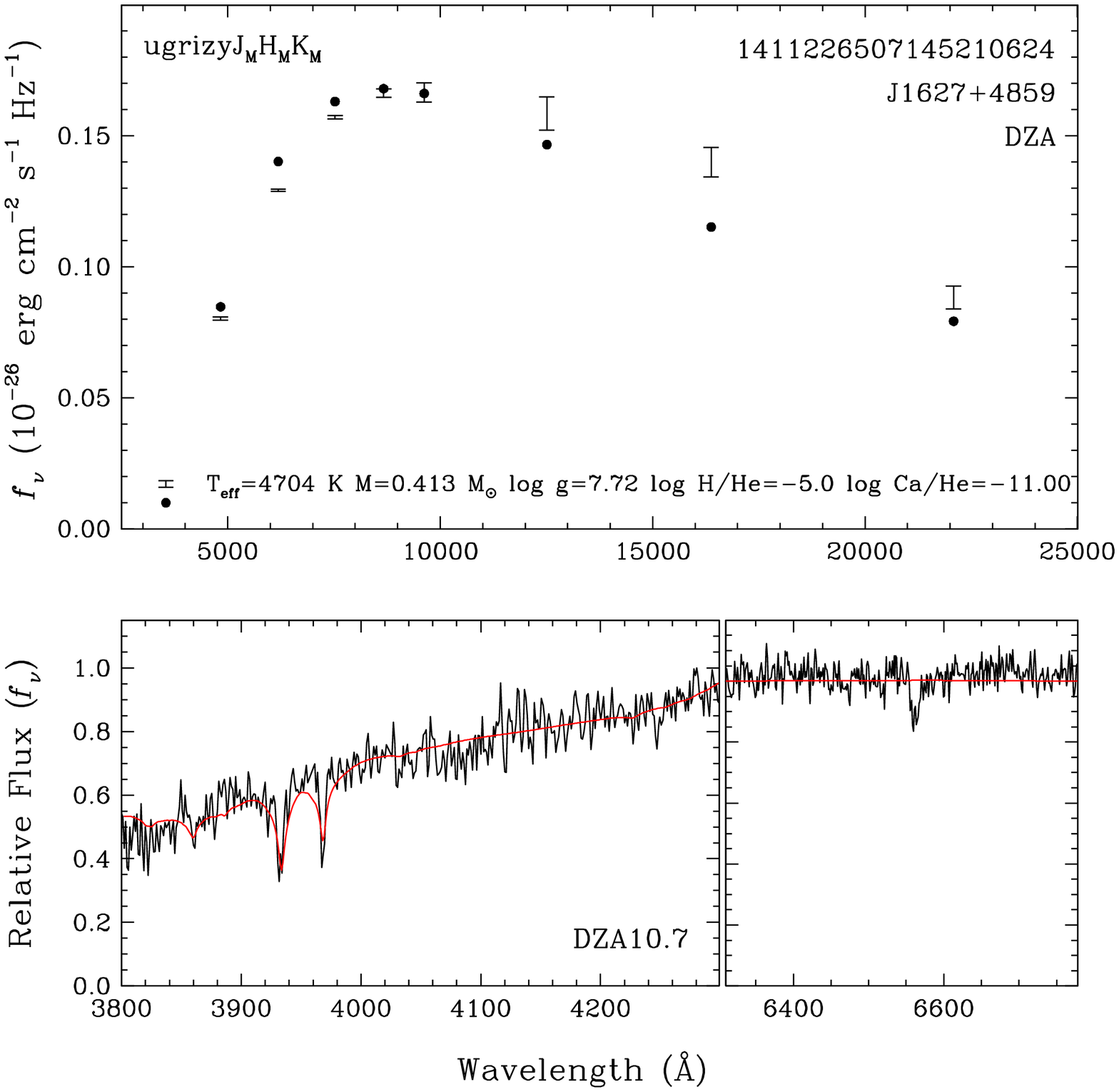}
\includegraphics[width=3.3in, clip=true, trim=0.0in 2.5in 0.0in 0.0in]{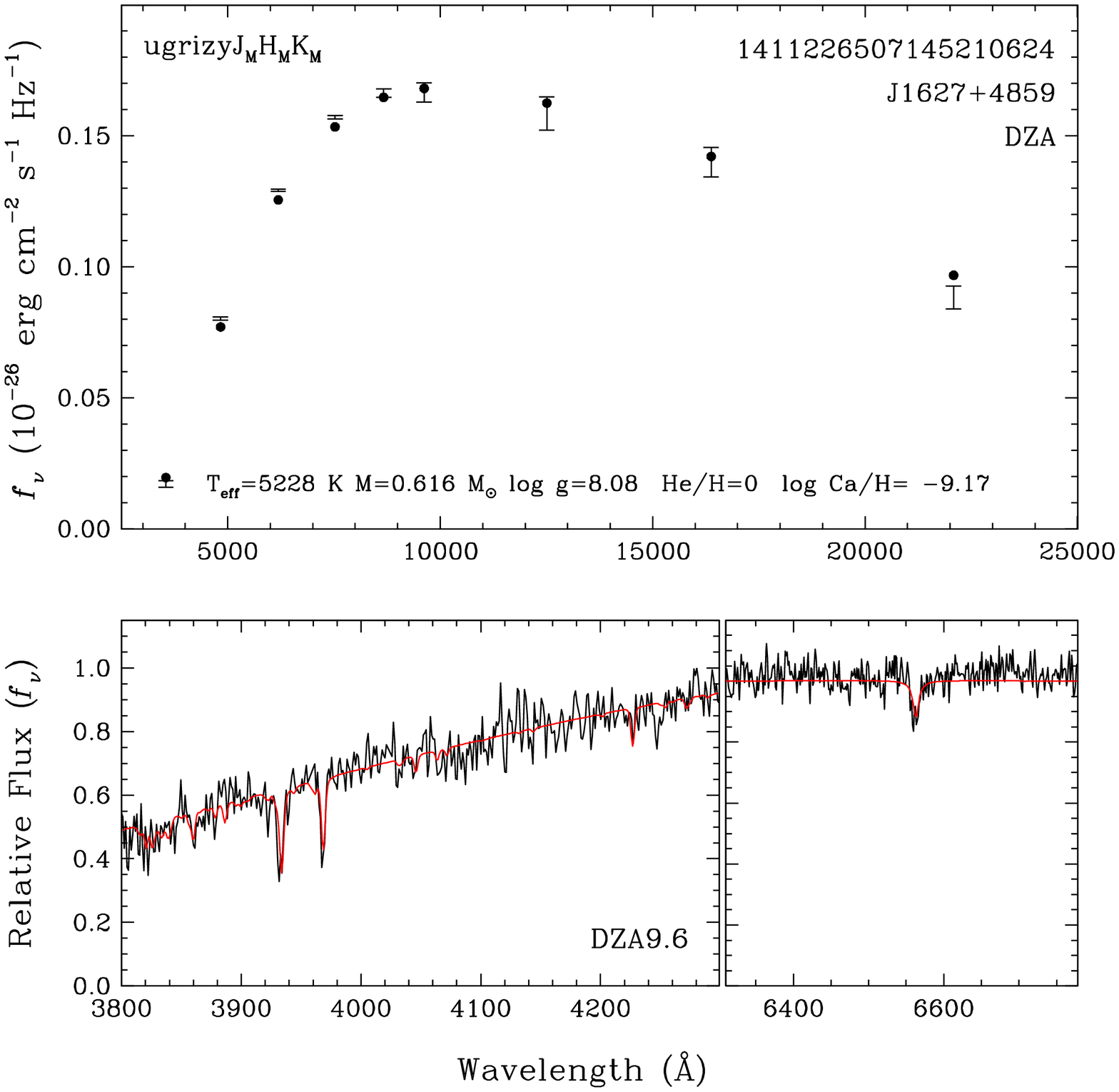}
\caption{Same as Figure \ref{fit_DZ} but for a DZA star fitted under the assumption
of a He-atmosphere (top two panels) and H-atmosphere (bottom two panels) white dwarf.
\label{fit_DZH}}
\end{figure}

An important issue in our analysis is whether the cool DZ and DZA
stars in our sample are H- or He-atmosphere white dwarfs (see, e.g.,
\citealt{blouin22}). To explore this possibility, we fitted all
relevant objects with the He-atmosphere DZ model grid described
above, and compared the solutions with those obtained from the model
grid for H-atmosphere DAZ white dwarfs described in Section
\ref{sec:DA}. We found out that it was rather easy to discriminate
both solutions by simply looking at the quality of the fits. But more
importantly, we also found that all cool, low-mass ($M<0.45$ \msun) DZ
and DZA stars uncovered using He-atmosphere models turned out to be
H-atmosphere white dwarfs, with much larger masses. A good example is
displayed in Figure \ref{fit_DZH} for the DZA white dwarf J1627+4859,
where we contrast both solutions. Here the mass obtained using
H-atmosphere models, $M=0.616$ \msun, is significantly larger than
that obtained using He-atmosphere models, $M=0.413$ \msun. The overall
quality of the fit is also far superior using H-atmosphere models.

\subsection{IR-faint White Dwarfs}\label{sec:IRF}

\citet{bergeron22} have recently demonstrated that white dwarfs showing
a strong infrared flux deficiency resulting from collision-induced
absorption (CIA) by molecular hydrogen -- often referred to as
ultracool white dwarfs -- are not so cool after all. Indeed, model
atmospheres including the high-density correction to the He$^-$
free-free absorption coefficient of \citet{iglesias02} have been
shown to yield significantly higher effective temperatures and larger
stellar masses than previously reported. Hence, these objects should be
more accurately referred to as IR-faint white dwarfs.

\begin{figure}
\centering
\includegraphics[width=3.3in, clip=true, trim=0.5in 3.in 0.8in 0.0in]{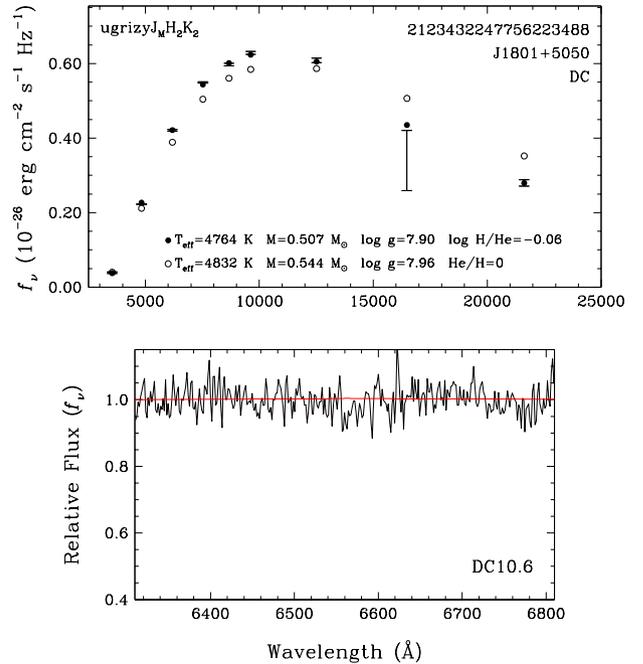}
\caption{Best fit to a new IR-faint white dwarf identified in our
  sample. Upper panel: The observed photometry, indicated at the top
  left, is shown by error bars, while the model atmosphere fit is
  shown by solid dots for the mixed H/He solution, and open circles
  for the pure H solution, with parameters given at the bottom. Lower
  panel: Optical spectrum (black) compared with the model spectrum
  (red) interpolated at the parameters obtained from the pure H photometric
  solution.
\label{fit_IRF}}
\end{figure}

We have 47 IR-faint white dwarfs in our sample, 43 of which have been
previously analysed in \citet{bergeron22}.  With the exception of
J0041$-$2221 (LHS 1126), which we analyzed above as a DQ white dwarf,
we fit the IR-faint white dwarfs in our sample using the same models
as those described at length in Bergeron et al.  These are all DC
white dwarfs with the exception of the four previously known DZ stars
J0804+2239, J1824+1212, J1922+0233, and J2317+1830
\citep{blouin18b,tremblay20,hollands21}. The four new
IR-faint white dwarfs discovered in our analysis are J0223+2055,
J0910+2554, J1801+5050, and J2347+0304. Three of these have comparable
temperatures ($\Te\sim4500$~K) and stellar masses ($M\sim 0.5$ \msun),
but more importantly, comparable chemical compositions in the range
$\logh\sim-1.0$ to 0.1 (i.e., nearly equal amounts of H and He
compared to other IR-faint white dwarfs). Our best fit to J1801+5050
is displayed in Figure \ref{fit_IRF}. These four new discoveries had
not been identified as IR-faint candidates in Bergeron et al.~because
their colours in the Pan-STARRS $M_g$ versus $(g-z)$ colour-magnitude
diagram put them near the bulk of white dwarfs along the faint end of
the main white dwarf sequence, making them impossible to identify
without proper near-infrared photometry. We come back to this
important point in Section \ref{sec:bottom}.

\subsection{DC White Dwarfs}\label{sec:DC}

\begin{figure}
\centering
\includegraphics[width=3.3in, clip=true, trim=0.5in 3.in 0.8in 0.0in]{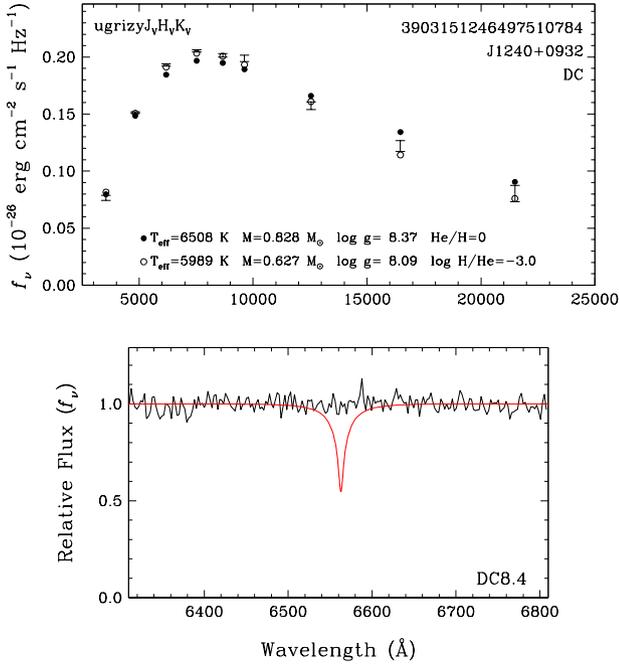}
\caption{Best fit to a typical DC white dwarf in our sample.  Upper
  panel: The observed photometry, indicated at the top left, is shown
  by error bars, while the model atmosphere fit is shown by solid dots
  for the pure H solution, and open circles for the mixed H/He
  solution, with parameters given at the bottom. Lower panel: Optical
  spectrum (black) compared with the model spectrum (red) interpolated
  at the parameters obtained from the pure H photometric solution.
\label{fit_DC}}
\end{figure}

The procedure used to fit the 810 DC white dwarfs in our sample
(excluding the IR-faint DC stars) is identical to the approach used
above for the DA stars (including the model atmospheres). A fit to a
typical DC star in our sample, J1240+0932, is displayed in Figure
\ref{fit_DC}. So far, it was possible for the previous spectral types
to obtain an almost unique solution because the information contained
in the optical spectrum (or the {\it JHK} photometry in the case of
IR-faint white dwarfs) allowed us to constrain the chemical
composition, whether it is the main constituent, hydrogen or helium,
or traces of other elements (hydrogen, carbon, metals).

For genuine DC stars with featureless spectra, it becomes extremely
difficult to determine even the main constituent of the atmosphere,
unless the object is hot enough that \ha\ would be detected if the
atmosphere were composed of hydrogen. The case of J1240+0932 displayed
in Figure \ref{fit_DC} is such an example.  Moreover, as demonstrated
in \citet{bergeron19}, invisible traces of hydrogen in He-atmosphere
DC white dwarfs can affect the temperature and mass determinations
significantly when using the photometric technique (see also Figure~5 
of \citealt{blouin19b}), and the exact
amount of hydrogen present is impossible to determine accurately.  The only
exception is for IR-faint DC white dwarfs where the presence of
hydrogen can be inferred, and thus measured, from the strong infrared
absorption resulting from collision-induced absorption by molecular
hydrogen, whether it is due to H$_2$--H$_2$ or H$_2$--He interactions,
but this becomes only possible at low effective temperatures
($\Te\lesssim6000$~K) when molecular hydrogen starts to form.

In principle, it is theoretically possible to distinguish H-rich from
He-rich white dwarfs over a wide range of effective
temperatures due to the different behaviour of H and He continuum
opacities, but this requires extremely precise photometry, and more
importantly, a very accurate magnitude-to-flux conversion, which is
not always the case, even for Pan-STARRS and SDSS magnitudes based on
the AB photometric system, as discussed by \citet{bergeron19}. The
problem becomes even more severe when multiple photometric
systems are combined, for instance optical and infrared photometry. Consequently,
it is extremely dangerous to allow the H/He abundance ratio to be a
free parameter for fitting DC white dwarfs, as emphasized in
\citet{blouin19b}. We propose a different statistical approach below.

\subsubsection{The Bottom of the Cooling Sequence}\label{sec:bottom}

We first investigate in this section the nature of the cool DC white
dwarfs at the faint end of the cooling sequence in the
colour-magnitude diagrams displayed in Figures \ref{color_mag_pan} and
\ref{color_mag_sdss}. In both diagrams, the 0.6 \msun\ CO-core models
with pure H (solid lines) and pure He (dotted lines) atmospheres fail
to match the faint sequence of DC white dwarfs, although the pure H
models are definitely much closer than the pure He models. Even though
the mixed $\logh=-5$ models (dash-dotted lines) match the DC sequence
rather well in the SDSS $M_g$ versus $(u-g)$ colour-magnitude diagram,
they fail miserably in the Pan-STARRS $M_g$ versus $(g-z)$ diagram due
to the onset of CIA. We also notice that there are very few objects
observed along the pure He sequence at faint luminosities, suggesting
that most cool DC white dwarfs have H-dominated atmospheres, at least
based on these diagrams. Furthermore, we can see in Figure
\ref{color_mag_pan} that the pure hydrogen sequence provides an
excellent match to the DA stars at higher temperatures, and that the
departure occurs below roughly $\Te\sim5300$~K, when \ha\ becomes
difficult to detect. But from an empirical point of view, the DA
sequence at this temperature merges perfectly well with the cool DC
sequence. All these observations strongly suggest a problem with the
physics of pure-H model atmospheres at very low effective
temperatures.

\begin{figure}
\centering
\includegraphics[width=3.3in, clip=true, trim=0.5in 6.5in 0.8in 0.0in]{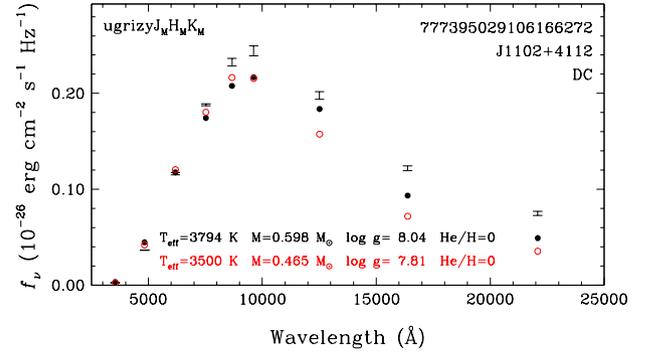}
\caption{Best fits to one of the coolest DC white dwarfs in our sample
  assuming a pure H composition. The observed photometry,
  indicated at the top left, is shown by error bars. The filled black
  dots show the fit obtained by including the near-infrared {\it JHK} photometry,
  while the open red dots show the solution obtained by fitting only the optical
  {\it ugrizy} photometry.  The corresponding parameters are given at the
  bottom.
\label{fit_coolestDC}}
\end{figure}

We investigated this problem by looking at one of the coolest objects
along the DC sequence, J1102+4112, with $M_g=17.28$ and $(g-z)= 2.01$
(the reddest DC star in Figure \ref{color_mag_pan}). This object has
also been analysed in detail by \citet{hall08}. Our best photometric
fit under the assumption of a pure H atmosphere is displayed in Figure
\ref{fit_coolestDC}. Here we show two fits, one by including the
near-infrared {\it JHK} photometry (the solution shown in black), and
the other solution (in red) where the {\it JHK} photometry has been
omitted. Although both fits are far from perfect, they are
significantly superior to those obtained with mixed H/He or pure He
atmospheres (not shown here), two solutions that can be easily ruled
out. Similar conclusions have been reached by \citet[][see their
  Figures 4 and 5]{hall08}.  The results shown in Figure
\ref{fit_coolestDC} also indicate that the solution based solely on
optical {\it ugrizy} photometry yields a low temperature of $\Te=3500$~K,
and a low mass of $M=0.465$~\msun. However, the
inclusion of the near-infrared {\it JHK} photometry has the effect of
pushing the solution towards a higher temperature of $\Te=3800$~K, and
a much larger mass of $M=0.598$~\msun\ -- i.e., $\sim$0.13 \msun\ larger --
in much better agreement with the average mass of white
dwarfs.\footnote{Note that \citet{leggett18} erroneously obtained a
  mixed H/He solution for this object (WD 1059+415 in their Table 4),
  although they also show (see the supplementary material) their best
  fit assuming a pure H composition, which is almost identical to that
  shown here. Their pure H solution fit even provides an excellent
  match to the Spitzer photometry.}

\begin{figure}
\centering
\includegraphics[width=3.3in, clip=true, trim=0.in 2.1in 0.0in 0.1in]{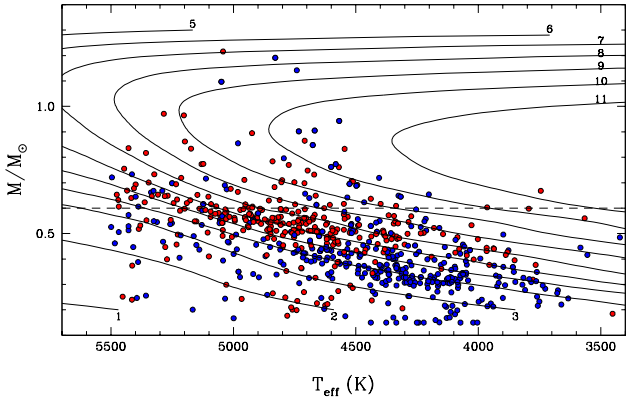}
\caption{Mass as a function of effective temperature for the cool DC
  white dwarfs in our sample under the assumption of pure H (red dots)
  and pure He (blue dots) atmospheric compositions. The dashed line
  corresponds to 0.6 \msun, the average mass for white dwarfs.  Solid
  curves are theoretical isochrones (white dwarf cooling age only),
  labeled in units of $10^9$ yr, obtained from the CO-core cooling sequences
  described in Section \ref{sec:CMD} with thick H layers ($\qh=10^{-4}$).
\label{MT_coolDC}}
\end{figure}

We observe a similar effect in all cool DC white dwarfs in our sample,
but to a lesser extent as we go up in temperature. And again, the pure
He model atmospheres yield far worse solutions, in particular in terms
of mass. This can be appreciated by examining the mass versus
effective temperature distribution for all the cool
($\Te\lesssim5500$~K) DC white dwarfs in our sample, displayed in
Figure \ref{MT_coolDC}, obtained under the assumption of pure H (red
symbols) and pure He (blue symbols) atmospheric compositions. Note
that for the objects below 0.2 \msun, we rely on the He-core models
described in \citet[][see their Figure 3]{bergeron22}, which are
appropriate in the temperature range considered here. We can see that
the $\Te$ and $M$ values obtained under the assumption of pure He
compositions are unrealistically low. This result is entirely
consistent with the prediction of the 0.6 \msun, pure He sequence in
the colour-magnitude diagram displayed in Figure \ref{color_mag_pan}
with respect to the location of these cool DC stars; the models need
to be more luminous, thus requiring larger radii and smaller
masses. In contrast, the masses obtained under the assumption of pure
H models in Figure \ref{MT_coolDC} are much closer to the mean mass of
white dwarfs, although we see a definite trend towards lower masses at
lower temperatures, another indication of problems with the physics of
pure H model atmospheres at low effective temperatures, and thus higher
photospheric densities, a conclusion also reached by \citet{hall08}.

We can now understand more quantitatively the results displayed in
Figure 12 of \citet{bergeron19}, which shows similar $M$ versus $\Te$
distributions for all white dwarfs found in Gaia within a distance of
100 pc, including both spectroscopically identified white dwarfs as
well as white dwarf candidates. In this case, the stellar parameters
have been determined using photometric fits to Pan-STARRS {\it grizy}
photometry only. Their solutions obtained from pure H (top panel) and
pure He (middle panel) model atmospheres show the same behavior as in
Figure \ref{MT_coolDC}, with the exception that our masses based on
pure H models are larger due to the use of {\it JHK} photometry for
many of our objects, but still not enough to reach the $\sim$0.6
\msun\ average mass for white dwarfs for the coolest DC stars in our
sample.

All the results above strongly support the idea, also suggested by
\citet{kowalski06}, that the bulk of cool ($\Te\lesssim5000$~K) DC
white dwarfs, but not all (see below), probably have H atmospheres. By
taking into account the suspected problems with the H-atmosphere
models, we did not find a single convincing case of a cool, pure He DC
white dwarf in our sample. Note that we exclude from the discussion
the cool DZ white dwarfs, which are discussed further below. The fact
that the 0.6 \msun\ pure H theoretical sequence in Figure
\ref{color_mag_pan} does not match the observed DC sequence is
obviously related to the less-than-perfect fit displayed in Figure
\ref{fit_coolestDC}, in particular in the optical. 

One possible explanation for this problem is the UV/blue opacity
originating from the red wing of Ly$\alpha$. \citet{kowalski06}
performed such calculations and successfully demonstrated that models
including this opacity show significant improvements in explaining the
location of DC white dwarfs in colour-colour diagrams (see their
Figure 4). Note that their calculations is based on the one-perturber
approximation, H or H$_2$, and it is possible that the interaction
from multiple perturbers (see, e.g., \citealt{allard09}) need to be
taken account in the coolest white dwarfs where the densities are
significantly larger.  We also note that \citet{saumon14} analysed
Hubble Space Telescope STIS spectra of eight very cool WDs (five DA
and three DC stars) and showed that their fits become worse at lower
$\Te$ and higher $\logg$ values.  As the authors conclude, these are
the stars with higher photospheric densities and more extreme physical
conditions, implying that the microphysics of dense matter probably
needs to be revised. 

It is also possible that other sources of opacity might be responsible
for the problems described above. For instance, the top panel of
Figure 17 from \citet{saumon22} shows that at the photosphere of a
pure H model at $\Te=4000$~K, the opacity from the red wing of
Ly$\alpha$ dominates in the UV, while the H$^-$ bound-free opacity and
the H$_2$-H$_2$ (and H$_2$-H) CIA opacity dominate in the optical and
near-infrared, respectively. Perhaps the physics from these last two
opacity sources needs te be revised as well.

\begin{figure}
\centering
\includegraphics[width=3.3in, clip=true, trim=0.0in 0.6in 0.6in 0.7in]{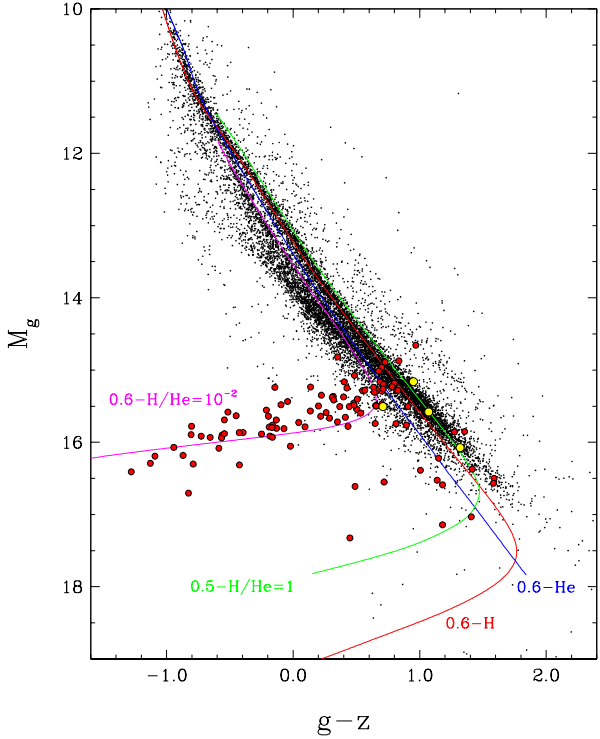}
\caption{Pan-STARRS $M_g$ versus $(g-z)$ colour-magnitude diagram for
  the 100 pc sample drawn from the MWDD. The red dots represent the
  IR-faint white dwarfs analysed in \citet{bergeron22}, while the
  yellow dots show the location of the four new objects identified in
  our analysis. Also overplotted are cooling sequences for 0.6
  \msun\ C/O-core models with pure H (red), pure He (blue), and
  $\logh=-2$ (magenta) atmospheric compositions.  The green sequence is
  for 0.5 \msun\ C/O-core models with $\logh=0$ (i.e., ${\rm H/He}=1$).
  \label{color_mag_IRF}}
\end{figure}

As discussed above, the end of the cooling sequence in Figure
\ref{color_mag_pan} is composed mainly of DC stars that are better
interpreted as H-atmosphere white dwarfs, but not all.  The glaring
exceptions are the IR-faint DC white dwarfs with mixed H/He
compositions. We reproduce in Figure \ref{color_mag_IRF} the $M_g$
versus $(g-z)$ colour-magnitude diagram from \citet{bergeron22}
together with the 105 IR-faint white dwarfs they analysed (in red) as
well as the four new objects (in yellow) identified in our survey and
discussed in Section \ref{sec:IRF}.  Most IR-faint white dwarfs in
this diagram are located on the nearly horizontal branch, consistent
with mixed H/He atmospheres with $\logh\sim-2$, a result confirmed
quantitatively by a more rigorous model atmosphere analysis (see
Figure 15 of \citealt{bergeron22}).  However, these appear more common
because they clearly stand out in such a colour-magnitude diagram, an
obvious selection effect. But there are other IR-faint white dwarfs
more difficult to detect, such as those found on the main branch at
the end of the cooling sequence. These include three of the four we
identified in our analysis, thanks to the available {\it JHK}
photometry, otherwise they would have gone unnoticed\footnote{These
  four IR-faint white dwarfs had not even been identified as potential
  candidates in the search for additional IR-faint candidates
  discussed in Section 5.4 of \citet{bergeron22}}. Note that the
parameters determined for three of the new IR-faint white dwarfs
identified in our survey ($M\sim 0.5$ \msun, ${\rm H/He}\sim 1$) are
entirely consistent with their location in the colour-magnitude
diagram, as illustrated in Figure \ref{color_mag_IRF}, where we
overplotted the corresponding cooling sequence (in green).  It is
therefore possible that many more are hiding in the same region of the
colour-magnitude diagram. This stresses the importance of eventually
securing {\it JHK} photometry for all spectroscopically confirmed
white dwarfs at the end of the cooling sequence.

\subsubsection{The Atmospheric Composition of DC White Dwarfs}\label{sec:DCcomp}

As discussed above, it is possible to obtain a unique solution for the
physical parameters of cool white dwarfs -- effective temperature,
stellar mass, atmospheric composition -- using the photometric
approach given that we have sufficient spectroscopic information
(hydrogen, carbon, and metallic absorption features). Even in the case
of IR-faint white dwarfs, the H/He abundance ratio can be determined
from the CIA opacity in the near and far infrared. However, for most
genuine featureless DC white dwarfs, the situation is nearly
hopeless. Indeed, as discussed in \citet{BRL97}, there are also subtle
differences between pure H and pure He spectral energy distributions
due to the presence (or absence) of the H$^-$ opacity in the infrared
(see their Figure 15), or to the onset of CIA absorption at low
temperatures (see, e.g., Figure 1 of \citealt{blouin19b}), but such
measurements can be extremely difficult and dangerous, in particular
given the precision and accuracy (or even availability) of infrared
photometric measurements.  We adopt here a different approach for the
DC white dwarfs in our sample.

Above some $\Te$ value, it is almost impossible to distinguish pure He
models from models including a trace of hydrogen. Yet, as shown by
\citet{bergeron19}, even undetectable traces of hydrogen can affect
the temperature and mass determinations significantly at all
temperatures of interest in our analysis. As a conservative estimate,
we adopt $\Te=6500$~K as the temperature above which all DC stars are
assumed to have mixed H/He compositions. We adopt the same approach as
in \citet[][see their Section 3.5]{kilic20} and adjust the H/He
abundance ratio as a function of $\Te$, from $\logh=-5$ at
$\Te=11,000$~K, up to $\logh=-3$ at $\Te=6000$~K, thus following the
predictions of the convective mixing scenario of
\citet{rolland18}. However, we also found out that assuming $\logh=-5$
throughout has little effect on the results (see below). It is of
course possible that some of these DC white dwarfs may contain no
hydrogen whatsoever, and thus their masses are probably underestimated
here, but there is no possible way to identify those in our sample,
neither spectroscopically, nor photometrically.

As discussed in the previous section, there is empirical evidence that
most, but probably not all, extremely cool DC white dwarfs have pure H
compositions. Here we adopt $\Te=5200$~K as the temperature below
which we assume a pure H atmospheric composition for the DC white
dwarfs in our sample. This corresponds also roughly to the temperature
above which it is possible to distinguish DA and DC stars from the
detection of \ha. It is of course possible to detect theoretically
\ha\ at much lower temperatures, a value that also depends on
$\logg$, but the spectral classification is particularly sensitive to
the signal-to-noise ratio of the observations.  For instance, we can
detect a weak \ha\ feature in the DA star J0700$+$3157 at
$\Te\sim4900$~K, but this is an extremely high S/N spectrum obtained
with the Hale 5 m reflector at Palomar Observatory by
\citet{greenstein86}, who contributed to reclassify many DC white
dwarf as DA stars. Similar results have been reported by
\citet{BRL97,BLR01}.

\begin{figure}
\centering
\includegraphics[width=3.3in, clip=true, trim=0.5in 3.in 0.8in 0.0in]{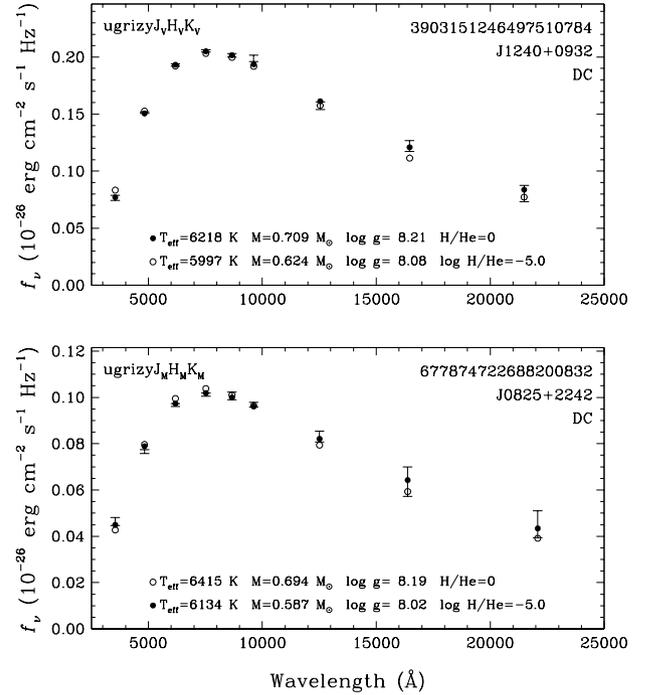}
\caption{Two DC stars in our sample that are better fit with pure He
  (top) or mixed H/He (bottom) model atmospheres. The observed
  photometry is shown as error bars and the photometric set used is
  indicated at the top left of each panel. Our best model fit,
  pure He or mixed H/He ($\logh=-5$), is shown as solid dots, while
  the other solution is shown as open circles with the model
  parameters given in each panel.
\label{fit_DCmixed}}
\end{figure}

Finally, there is a particular range in temperature between roughly
$\Te=5200$~K and 6500~K where the absence of \ha\ clearly rules out a
pure H composition, but where the spectral energy distribution remains
particularly sensitive to the presence of even small traces of
hydrogen due to the onset of the H$_2$ CIA opacity. We determined that
in this range of temperature, it is thus possible to differentiate
pure H from He-rich white dwarfs by the presence of \ha, and to
differentiate pure He from mixed H/He atmospheres by inspecting the
spectral energy distribution. For instance, we compare in Figure
\ref{fit_DCmixed} our best fits to two cool DC stars in this
temperature range, obtained with pure He and mixed H/He ($\logh=-5$)
compositions. J1240+0932 (top) is an example of a DC star that is
better fit under the assumption of a pure He atmosphere, while
J0825+2242 (bottom) is better fit by a mixed H/He atmosphere, a
conclusion that can be drawn by a simple inspection of the fits, but
also confirmed from our $\chi^2$ analysis. Note that the small trace
of hydrogen of only $\logh=-5$, which is much smaller than the
hydrogen abundances inferred in IR-faint white dwarfs (see Figure 15
of \citealt{bergeron22}), can affect both the temperature and mass
determinations significantly. Moreover, a comparison of the $\logh=-5$
solution for J1240+0932, displayed here, with the $\logh=-3$ solution
previously shown in Figure \ref{fit_DC} for the same object, indicates
that the particular H/He value used for DC stars in this temperature
range (and above) does not affect the parameter determinations
significantly.

To summarize, for temperatures between $\Te=5200$~K and 6500~K, we
adopt the pure H solution for DA stars, while for DC stars we adopt
the pure He or mixed H/He solution based on a simple $\chi^2$
analysis. We stress that this approach remains approximate because not
all white dwarfs have near-infrared photometry available, and even for
those that do, the quality of the photometry is highly variable.

\subsection{Magnetic White Dwarfs}\label{sec:mag}

\begin{figure}
\centering
\includegraphics[width=3.3in, clip=true, trim=0.5in 3.in 0.8in 0.0in]{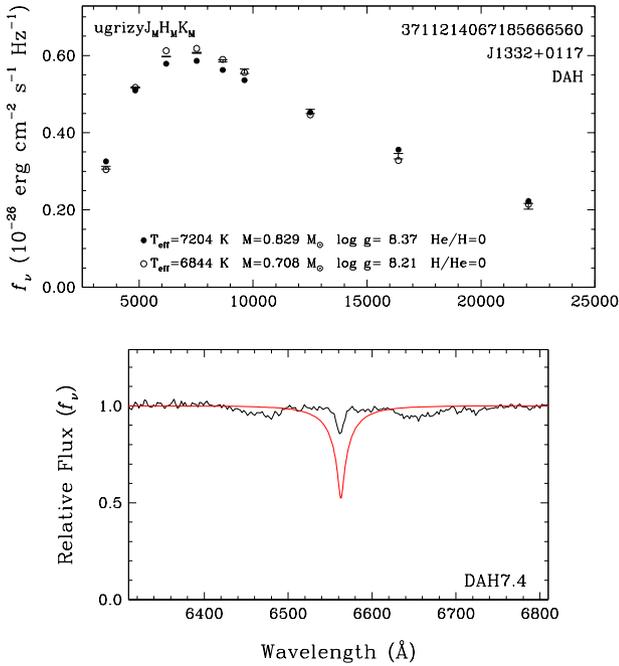}
\caption{Best fit to a magnetic DA star in our sample. The observed
  photometry is shown as error bars and the photometric set used is
  indicated at the top left of the upper panel. Our pure H and pure He
  model atmosphere fits are shown as solid and open circles,
  respectively, with the parameters given in the upper panel. The
  lower panel shows the normalized optical spectrum, compared with the
  model spectrum (in red) interpolated at the parameters obtained from
  the pure H photometric solution.
\label{fit_DAH}}
\end{figure}

We have several magnetic white dwarfs in our sample of various
spectral types. Among those, the more obvious are the magnetic DA
stars, with Zeeman splitting in the form of a triplet at \ha\ for a
moderate field (see, e.g., Figure \ref{fit_DAH}), or with more
displaced and distorted components if the magnetic field is much
stronger. We also have several magnetic DQ white dwarfs, although in this
case, these are restricted to the very cool DQ stars with strong and
shifted C$_2$ Swan bands (see, e.g., J1038$-$2040 $\equiv$ LP
790-29). We also have many cases of magnetic DZ stars (see, e.g.,
J1515+8230). In our analysis, we neglect the existence of these
magnetic fields and apply the photometric method using non-magnetic
model atmospheres. One obvious effect of using magnetic models is to
change the predicted photometry due to absorption lines that are
shifted to different wavelengths as a result of the magnetic
field. However, we also found cases where even when line blanketing is
unimportant, the continuum opacity seems to be affected, an example of
which is displayed Figure \ref{fit_DAH}, where we see that the energy
distribution is better reproduced by a pure He model rather than a
pure H model. We also observe a similar phenomenon in hotter magnetic
DA stars, where the Balmer jump appears to be suppressed, in
particular in objects with SDSS {\it u} magnitudes available (see,
e.g., J1126+0906). Despite the better agreement with pure He (or
He-rich) models for these stars, we still adopt the pure H solutions
for all magnetic DA white dwarfs in our sample.

\subsection{Unresolved Double Degenerate Binaries}\label{sec:DD}

The existence of unresolved double degenerate binaries (DD) in our
sample can already be inferred from the colour-magnitude diagram
displayed in Figure \ref{color_mag_pan} where they appear as
overluminous objects above the main white dwarf sequence. Indeed, two
unresolved white dwarfs of equal luminosity would appear $\sim$0.75
mag brighter than a single object, consistent with the location of
most overluminous objects in Figure \ref{color_mag_pan} (for systems
composed of two $\sim$0.6 \msun\ white dwarfs). We also note that most
spectroscopically identified overluminous white dwarfs in Figure
\ref{color_mag_pan} are DA and DC stars, although we see a few DQ
stars as well.

When applying the photometric technique to these DD systems assuming a
single star, the radius measurement is usually overestimated since two
objects contribute to the total luminosity, and the stellar mass is
consequently underestimated as well. However, this may not necessarily
be the case if one white dwarf component completely dominates the
overall luminosity of the system, in which case the radius and mass
measurements remain accurate. If an optical spectrum is available, the
spectroscopic technique can be combined with the photometric method to
deconvolve the stellar parameters of each component of the system, as
done for instance by \citet[][see their Figures 14 to 16]{bedard17}.

In our spectro-photometric analysis, we are able to detect the
presence of DD systems in various ways. The first is of course when
the stellar mass is very low ($M\lesssim 0.48$ \msun), because such
low-mass white dwarfs cannot have formed within the lifetime of our
Galaxy, and can therefore only exist as a result of common envelope
evolution. One example in our sample is the DA J0129+1022 with $M=0.340$
\msun. In this particular case, the predicted spectrum at \ha\ is in
perfect agreement with the measured parameters obtained from the fit
to the energy distribution assuming a single star. Hence, it is
possible here that the luminosity of the brighter component of the
system completely dominates that of the second
component. Alternatively, if two DA components have similar parameters
and contribute equally to the total luminosity of the system, we
would also reach a perfect agreement between the photometric and
spectroscopic solutions. It is of course impossible to distinguish
between these two possibilities without a more refined analysis.

In most cases, however, it is expected that the individual components
of DA + DA unresolved systems will have different stellar parameters,
in which case we may observe a disagreement between the observed
\ha\ profile and that predicted from the photometric solution assuming
a single star. For instance, J0532+0624 ($M=0.253$ \msun) shows a
distorted energy distribution that resembles a He-rich white dwarf,
and moreover, the predicted \ha\ profile is at odds with the
observations. Another example is J0138$-$0459 (L870-2, EG 11, WD
0135$-$052), a classical case of a DD system composed of two DA white
dwarfs where both components contribute to the total luminosity of the
system, but with significantly different parameters \citep{bergeron89}.

Even though all low mass white dwarfs must have formed through binary
evolution, the reverse is not necessarily true, and there may exist
more normal mass, or even massive, white dwarfs in our sample that are
hidden in DD systems. For instance, J0611+0544 ($M=0.594$ \msun),
J0904+1349 ($M=0.611$ \msun), and J0929$-$1732 ($M=0.632$ \msun) are
good examples of normal mass white dwarfs with observed
inconsistencies in there fits, while J0855$-$2637 ($M=0.840$ \msun) is
a good example of a massive object.

It is also easy to detect the presence of DD systems when the spectral
types of both components differ. A good example is J0110+2758
($M=0.301$ \msun) where the broad and shallow \ha\ feature suggests
the presence of a DC component in the system that dilutes the
\ha\ absorption profile. Similarly, we have 11 DQ stars with inferred
masses below 0.4 \msun\ (8 with $M<0.3$ \msun), although 4 of these
are DD systems composed of a DQ and a cool DA white dwarf (J0613+2050,
J0928+2638, J1406+3401, and J1532+1356). Perhaps the remaining 7
objects are also DD systems with the other component being a
featureless DC white dwarf. Interestingly, we find no low-mass
($M<0.4$ \msun) DZ star in our sample; the only one
is the IR-faint white dwarf J1824+1212 ($\Te\sim4700$~K).

In what follows, we make no attempt to deconvolve the individual
components of DD systems and simply assume that they are single white
dwarfs. We just have to keep this in mind when interpreting the
physical parameters obtained from our fits.

\subsection{White Dwarfs + Companions}\label{sec:mdwarf}

Because of the colour cut in the selection of our white dwarf sample, we avoid most WD + M--dwarf
binaries, but these systems still exist in our sample, in particular
if the M--dwarf has a late type that contaminates the colours, mostly in
the near-infrared. In such cases, the M--dwarf companion
may contribute significantly to the total flux,
which leads to an overestimation of the white dwarf radius, and to an
underestimation of its mass. Here we simply exclude the contaminated
bandpasses from our photometric fits (these excluded bandpasses are shown in red in our figures). A
good example is the DZ white dwarf J1651+6635. Contamination may also
occur from different spectral types such as brown dwarfs, like in the
case of the DA\,+\,L5 system J0135+1445 \citep[NLTT
  5306B,][]{casewell20}.

\subsection{Adopted Parameters}\label{sec:param}

The physical parameters for all white dwarfs in our sample are
presented in Table 2, where for each object we provide the J name,
Gaia ID (DR2; if not available the EDR3 is given and marked by a
  star symbol in Table 2), spectral type, effective temperature
($\Te$), mass ($M$/\msun), surface gravity ($\logg$), chemical
composition (see below), luminosity ($\log L$/\lsun), and white dwarf
cooling age ($\tau$ in Gyr).

\begin{table*}
\centering
\caption{Physical Parameters of the 100 pc White Dwarf Sample. This table is available in its entirety in machine-readable form.}
\begin{tabular}{crcccccccc}
\hline
\hline
Name & Gaia ID (DR2/EDR3*) & Sp Type & $\Te$ (K) & $M$/\msun & $\logg$ & Comp & Metals & $\log L$/\lsun & $\tau$ (Gyr)  \\
\hline

J0000+0132 & 2738626591386423424 & DA & 10342 (32) & 0.622 (0.007) & 8.034 (0.008) & He/H=0 & $\cdots$ & $-$2.79 & 0.61\\
J0000+1906 & 2774195552027050880 & DC & 5069 (54) & 0.553 (0.055) & 7.955 (0.066) & He/H=0 & $\cdots$ & $-$4.00 & 4.86\\
J0001+3237 & 2874216647336589568 & DC & 5753 (56) & 0.560 (0.034) & 7.981 (0.041) & H/He=0 & $\cdots$ & $-$3.80 & 3.06\\
J0001+3559 & 2877080497170502144 & DC & 6177 (73) & 0.688 (0.031) & 8.184 (0.035) & H/He=0 & $\cdots$ & $-$3.79 & 3.49\\
J0001$-$1111 & 2422442334689173376 & DC & 6541 (48) & 0.651 (0.024) & 8.124 (0.027) & log H/He=$-$3.2 & $\cdots$ & $-$3.65 & 2.41\\
J0002+0733 & 2745919102257342976 & DA & 7727 (44) & 0.598 (0.015) & 8.007 (0.018) & He/H=0 & $\cdots$ & $-$3.29 & 1.23\\
J0002+0733 & 2745919106553695616 & DAH & 8130 (71) & 0.805 (0.017) & 8.331 (0.02) & He/H=0 & $\cdots$ & $-$3.39 & 1.90\\
J0002+1610 & 2772241822943618176 & DA & 6684 (42) & 0.613 (0.028) & 8.037 (0.033) & He/H=0 & $\cdots$ & $-$3.56 & 1.87\\
J0002+6357 & 431635455820288128 & DC & 4959 (23) & 0.596 (0.01) & 8.026 (0.011) & He/H=0 & $\cdots$ & $-$4.08 & 6.55\\
J0003+6512 & 432177373309335424 & DC & 8595 (95) & 0.506 (0.014) & 7.874 (0.018) & log H/He=$-$4.0 & $\cdots$ & $-$3.04 & 0.82\\

\hline
\end{tabular}
\end{table*}

For clarity, we briefly summarize the procedure used to measure these
parameters, and in particular the chemical composition. For DA stars,
we assume a pure hydrogen composition (${\rm He/H}=0$), with the
exception of He-rich DA stars for which the H/He ratio ($\logh$) is
adjusted to match the \ha\ absorption profile.  For DQ stars, we
provide the carbon abundance ($\log {\rm C/He}$) measured using the
optical spectrum. For DZ stars -- including DZA and DAZ stars as well
--, we give the calcium abundance relative to helium ($\log {\rm
  Ca/He}$) or hydrogen ($\log {\rm Ca/H}$) depending on the main
atmospheric constituent, measured using the metallic features observed
in the optical spectrum (all other metals have chondritic abundance
ratios with respect to Ca). For IR-faint white dwarfs, we provide the
H/He ratio ($\logh$) obtained from fits to the energy
distribution. Finally, for DC stars, we assume a pure hydrogen
composition (${\rm He/H}=0$) for objects below $\Te=5200$~K; between
$\Te=5200$~K and 6500~K, we adopt the pure He (${\rm H/He}=0$) or
mixed H/He solution ($\logh$) based on a $\chi^2$ analysis; for
$\Te>6500$~K, we also adopt a mixed H/He solution where the hydrogen
abundance is adjusted as a function of $\Te$ (as discussed above, the
precise hydrogen abundance has little effect on the measured
parameters).

\section{Selected Results}\label{sec:results}

We can now explore the global properties of our sample, which is
composed of all white dwarfs identified spectroscopically in the MWDD
and with effective temperatures below $\Te\sim10,000$~K. As such, it
is by no means statistically complete in any sense, although we can
expect with such a large sample that all spectral types are
statistically well sampled at a given temperature, but probably not as
a whole. For instance, \citet{kilic20} increased significantly the
number of spectroscopically confirmed white dwarfs within 100 pc in
the SDSS footprint, but their sample was concentrated above $\sim$6000
K where DA stars could be easily identified. Hence, important
selection effects are present in our sample, and we make no attempt in
our analysis to restrict the results to any particular volume-limited
sample.

\subsection{Mass Distributions}\label{sec:mdistr}

\begin{figure*}
\centering
\includegraphics[width=4.5in, angle=270, clip=true, trim=0.in 0.in 0.0in 0.0in]{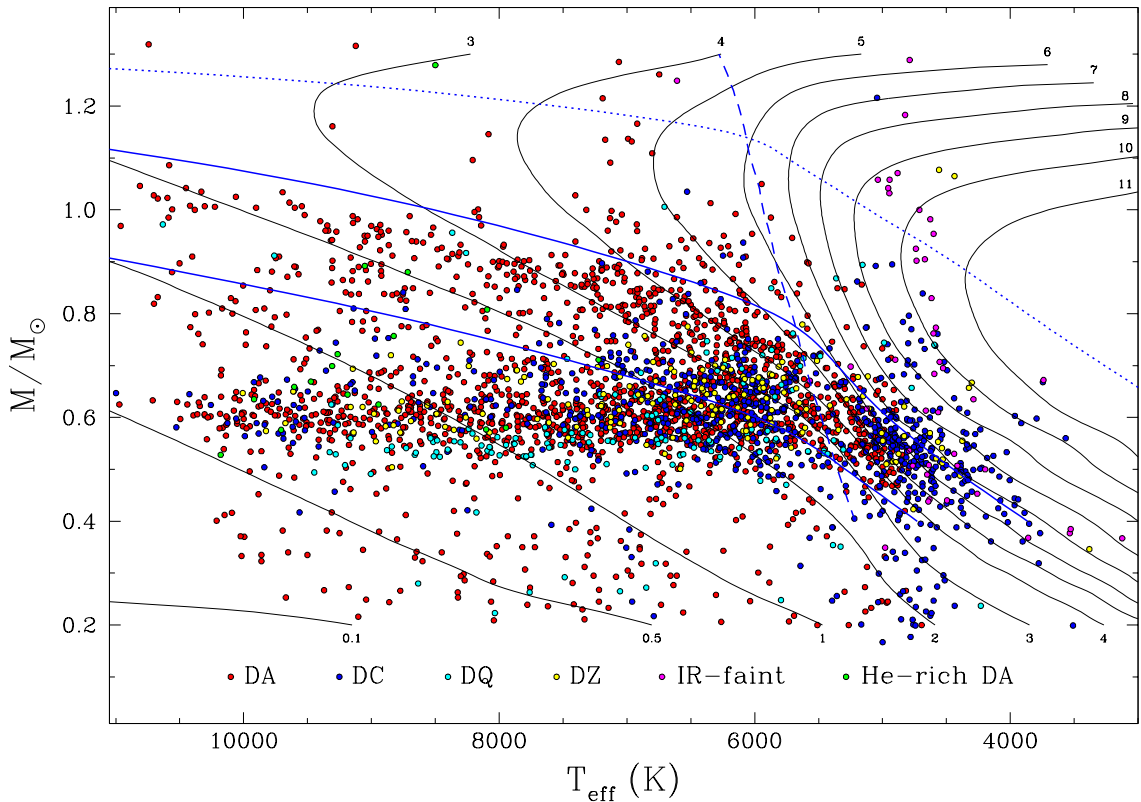}
\caption{Mass distribution as a function of effective temperature for
  all 2880 spectroscopically confirmed white dwarfs in our sample for
  the various spectral types indicated in the legend. Solid black
  curves are theoretical isochrones (WD cooling age) labeled in units
  of Gyr, obtained from cooling sequences with C/O-cores, $q({\rm
    He})=10^{-2}$ and $q({\rm H})=10^{-4}$ (see Section
  \ref{sec:CMD}). The lower blue solid curve indicates the onset of
  crystallization at the centre of evolving models, and the upper one
  indicates the locations where 80\% of the total mass has
  solidified. The dashed curve indicates the onset of convective
  coupling, while the dotted curve corresponds to the transition
  between the classical to the quantum regime in the ionic plasma (see
  text).
\label{mass_teff}}
\end{figure*}

The mass distribution, $M$ versus $\Te$, for all 2880
spectroscopically confirmed white dwarfs in our sample are displayed
in Figure \ref{mass_teff}, where the various spectral types are
indicated with different colours.  Also superposed are theoretical
isochrones (WD cooling age only) obtained from cooling sequences with
C/O-core compositions, $q({\rm He})=10^{-2}$ and $q({\rm H})=10^{-4}$;
the other curves are described below. In terms of the overall analysis
and model atmospheres, this figure is very similar to that shown in
Figure 18 of \citet{kilic20}, with the exception that our analysis is
not restricted to the SDSS footprint, it makes use of extensive sets
of near-infrared {\it JHK} photometry, and more importantly, it is not
restricted in temperature at the cool end of the white dwarf sequence
(Kilic et al.~restricted their analysis to $\Te\gtrsim 6000$~K). There
are many interesting features observed in Figure \ref{mass_teff},
which we now discuss in turn.

We first focus on the region above $\Te\sim6000$~K where we can see a
nice clustering of objects around the mean mass for white dwarfs
($M\sim0.6$ \msun) composed mostly of DA stars, but also a significant
fraction of non-DA stars including DC, DZ, and He-rich DA white
dwarfs. Note that the normal masses obtained here for the DC stars can
only be achieved with model atmospheres containing traces of hydrogen
($\logh\sim-5$). The same conclusion applies to DZ white dwarfs, but
in this case metals are the main source of free electrons. The DQ
stars in the same temperature region also form a tight sequence, but
with a mean mass $\sim$0.05 \msun\ lower than average, consistent with
the results obtained by \citet[][see their Figure 13]{coutu19}. There
are at least two possible explanations for this lower mean mass, one
involving problems with the physics of DQ model atmospheres
\citep{coutu19}, and another one recently proposed by
\citet{bedard22b}, who suggested that carbon (and hence the DQ
phenomenon) is preferentially detected in lower mass white dwarfs.

There are also two other regions where DA stars are concentrated. The
first one is along the crystallization sequence at high mass. To
illustrate this process, we indicated by the lower solid blue curve in
Figure \ref{mass_teff} the region where crystallization starts at the
centre of the white dwarf, and by the upper solid blue curve the limit
of 80\% crystallization of the core resulting from the solidification
front moving upward in the star with further cooling. When
crystallization occurs, the release of latent heat and chemical
fractionation decrease the cooling rate of a white dwarf, leading to
this obvious pile up of objects in the $M$ versus $\Te$ diagram
observed in Figure \ref{mass_teff}. Note that the hot end of the
crystallization sequence also contains several massive DQ, DC, and
He-rich DA white dwarfs. Furthermore, we can clearly see the
crystallization sequence merging with the normal-mass white dwarf
sequence near $\Te\sim5500$~K. Incidentally, this also coincides with
the temperature where convective coupling occurs at $M\sim0.6$ \msun,
that is, when the convection zone first reaches into the degenerate interior
where all of the thermal energy of the star resides. This is illustrated by
the dashed curve in Figure \ref{mass_teff}, which marks the effective
temperature of the onset of convective coupling as a function of mass.

DA stars are also found in large numbers at very low masses
($M\lesssim 0.45$ \msun). These are most likely unresolved DD
binaries, and the low masses inferred here are simply the result of
these objects being overluminous, as observed in Figure
\ref{color_mag_pan}. For instance, we can calculate that two identical
0.6 \msun\ unresolved DA white dwarfs would result in a $\sim$0.32
\msun\ single object in this $M$ versus $\Te$ diagram. But as
discussed above, it is also possible that some of these low mass
determinations are real if one component of the DD system completely
dominates the overall luminosity. Not all low-mass white dwarfs are DA
stars, however, and we can see several DQ and DC white dwarfs as
well. Not surprisingly, the four DA + DQ unresolved binaries in our
sample are found at such low masses. Actually, all four systems have
inferred masses well below 0.3 \msun\ (the DQ stars J0941+0901,
J1135+5724, J1501+2100, and J1803+2320 also have masses below 0.3
\msun); if we assume that both the DA and DQ stars in these systems
contribute equally to the luminosity, we can calculate the mass of
the DQ components to be around 0.52 \msun, in excellent agreement with
the bulk of single DQ white dwarfs in this temperature range. It is
therefore possible that the remaining low-mass DQ white dwarfs in our
sample are also DD systems composed of a DQ white dwarf and a
featureless DC component. We come back to this point further
below. Similarly, it is also possible that the low-mass DC white
dwarfs above 6600 K are unresolved binaries, but composed of two normal-mass DC
stars, a result that would be consistent with the absence of low-mass
DB white dwarfs found in the spectroscopic analysis of
\citet{genest19b}.

We now turn our attention to the region below $\Te\sim6000$~K, where
the interpretation of our results becomes significantly more
complicated. The most obvious feature in this temperature range is the
trend of DA and DC white dwarfs towards lowers masses, reaching
$M\sim0.4$ \msun\ at the end of the cooling sequence. We remind the
reader that we assumed pure H atmospheres for these DC
stars. Had we used pure He or mixed H/He compositions, the
situation would have been even worse (see Figure \ref{MT_coolDC}). We
suggested that improved model atmospheres -- most likely related to
some inaccurate opacity source -- could yield larger masses and higher
effective temperatures for these cool DC white dwarfs, and cool DA
stars as well, thus filling the apparent gap in the white dwarf
distribution observed around $\Te\sim5300$~K. Interestingly enough, we
can also see that in the same temperature range, several DZ stars
appear to follow the $\sim$0.6 \msun\ sequence nicely, in contrast
with the cool DA and DC white dwarfs. In such cool DZ stars, the
presence of metals reduces the atmospheric pressure significantly, and
thus the effect of collision-induced absorption by molecular hydrogen,
if present. In fact, we found that traces of hydrogen have little
effect on the measured stellar parameters in this temperature range.
Also, we originally identified several low-mass DZ white dwarfs in
this temperature range, but it turns out these were H-dominated rather
than He-dominated DZ white dwarfs, as discussed in Section
\ref{sec:DZ}. Given these results, we also
explored the possibility that the coolest DC stars in our sample have
He-dominated atmospheres containing invisible traces of metals that
could affect their temperature and mass determinations. However, the
amount of metals required to affect the stellar parameters are large
enough that the star would appear as a DZ white dwarf instead of a DC
star, a conclusion previsouly reached by \cite{blouin18b}. 
The fact that DZ white dwarfs behave normally at low
temperatures again suggests that the low masses obtained for the cool
DA and DC white dwarfs in our sample are most likely related to
problems with the physics of pure-H model atmospheres.

Figure \ref{mass_teff} includes 47 IR-faint white dwarfs, which all
show strong collision-induced absorption by molecular hydrogen in the
near-infrared, 43 of which are in common with the analysis of
\citet{bergeron22}. Many of these appear to define the cool edge of
our sample at all masses, and in particular in the upper right corner
of the $M$ versus $\Te$ diagram. The two DZ stars in the same region
of the diagram are J1922+0233 and J2317+1830, also IR-faint white
dwarfs (see Section \ref{sec:IRF}). There is also one massive DC star
overlapping this region, J1305+7022. Unfortunately, only Pan-STARRS
{\it grizy} photometry is available, but our excellent fits with pure
H or pure He models suggest that this is probably not an IR-faint
white dwarf. As discussed at length in \citet{bergeron22}, the coolest
and most massive IR-faint white dwarfs in Figure \ref{mass_teff} are
found in the Debye cooling phase, when the specific heat decreases
quickly with cooling, thus rapidly depleting the reservoir of thermal
energy and producing the extreme increase of the cooling rate observed
in the upper right corner of Figure \ref{mass_teff}. To better
illustrate this phenomenon, we indicated by a dotted curve the
so-called Debye cooling phase, i.e., the rapidly cooling phase
resulting from the transition, in the solid phase, from the classical
regime where the specific heat of a solid is independent of
temperature, to the quantum regime where it goes down from that
constant value with decreasing temperature. As described in
\citet{bergeron22}, we define this transition from the classical to
the quantum regime by isolating the evolving model where the central
temperature becomes equal to 1/20 the central Debye temperature
($\theta_D/T=20$), as defined in \citet{althaus07}. Only a dozen or so
objects in our sample have reached the Debye cooling phase.  We also
emphasize that the presence of cool, IR-faint white dwarfs in this
particular cooling phase is the direct result of the improved model
atmospheres described in \citet{bergeron22}, which led to higher $\Te$
and $M$ values.

\subsection{Low-Mass White Dwarfs}\label{sec:lowmass}

We now focus our attention towards the low-mass ($M<0.45$ \msun)
white dwarfs in Figure \ref{mass_teff}, which include mostly DA stars,
but also several DQ, DZ, and DC white dwarfs. We consider each of the
non-DA spectral types in turn.

One of the DZ stars is J1824+1212, also an IR-faint white dwarf with
an extremely low temperature of $\Te=3381$~K (and $M=0.346$~\msun),
where the physics of our model atmospheres is most uncertain. The other DZ star
is J0045+1420, a problematic object in our sample, probably magnetic,
with an energy distribution that is not well reproduced by our
models. The low masses inferred for these two DZ white dwarfs are thus
uncertain.

\begin{figure}
\centering
\includegraphics[width=3.2in, clip=true, trim=0.in 4.0in 0.0in 0.in]{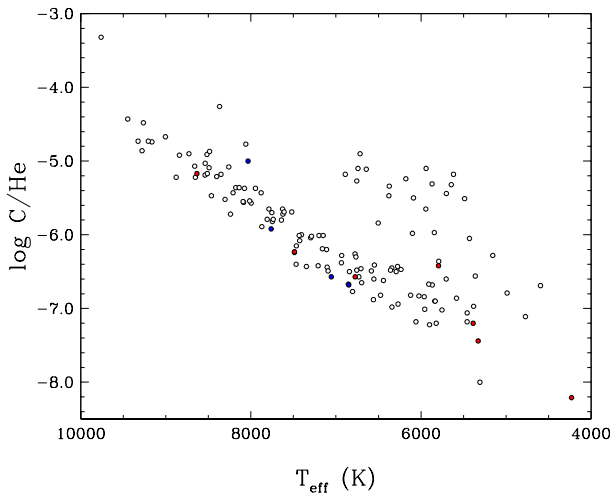}
\caption{Carbon abundance as a function of effective temperature for all DQ white dwarfs in our sample.
Coloured symbols all have $M<0.45$ \msun, with the blue dots representing the four unresolved DA + DQ
double degenerate systems. 
\label{compDQ}}
\end{figure}

Among the low-mass DQ white dwarfs, we already discussed in Section
\ref{sec:DQ} the four DA + DQ unresolved binaries in our sample. To go
further, we show in Figure \ref{compDQ} the carbon abundance as a
function of effective temperature for all DQ white dwarfs in our
sample, with these four binaries displayed in blue (all of these have
$M<0.3$ \msun), while the remaining DQs with $M<0.45$ \msun\ are shown
in red. Three of these binaries have C abundances near the bottom of
the envelope defining the DQ sequence, as expected if there is a
companion that dilutes the strength of the carbon Swan bands,
resulting in C abundances being underestimated (the other DA + DQ
system above the sequence is J0613+2050, an extreme case of
contamination by a DA component). We note that other low-mass DQ stars
in our sample have carbon abundances near the bottom of the envelope
(red symbols in Figure~\ref{compDQ}),
suggesting that these may also be unresolved double degenerates, but
this time composed of a DQ white dwarf and a featureless DC
component. The only exception is J1501+2100 ($M=0.248$~\msun), with a
large carbon abundance above the sequence. Perhaps in this case there is
indeed a low-mass DQ white dwarf in the system, dominating the
luminosity, with a C abundance higher than average, which is expected
in low-mass DQ white dwarfs, as discussed in \citet{bedard22b}. Given
these results, we inspected closely our fits to all low-mass DC white
dwarfs and discovered that J0822+2023 ($M=0.349$~\msun) shows a weak
absorption band near 5150 \AA, indicating that this is probably an
unresolved DQ + DC system as well.

With little evidence for low-mass single DQ white dwarfs, we
examined the remaining low-mass ($M<0.45$ \msun) DC stars in our
sample, but this time by restricting our inspection to $\Te>5500$~K
where the photometric analysis is less problematic. We find only 11 DC
stars, three of which have no spectra available to us, and the spectral
classification is thus difficult to confirm. A few of the remaining
DCs have M--dwarf companions with bad photometric fits (see, e.g.,
J0200+1222 with $M=0.217$~\msun), while in some cases, we can even
detect a hint of H$\alpha$ (see J1325+4523), although the spectrum is
rather noisy. Similarly, J1047+0007 ($M=0.245$~\msun) appears to show
Zeeman splitting at H$\alpha$ and could be magnetic. Hence, although
we cannot completely rule out the existence of low-mass He-atmosphere
white dwarfs in our sample, evidence suggests that most, if not all
low-mass objects in our sample have hydrogen atmospheres. This result
agrees with the conclusions of \citet{genest19b} who showed that there
is no evidence for low-mass single DB white dwarfs, as determined from
spectroscopy; several DB + DB systems have been identified by
photometry, however, but these are probably composed of normal mass DB white
dwarfs.

\subsection{The Evolution of DA White Dwarfs}\label{sec:evolda}

According to the analysis of \citet{bedard20}, about 75\% of all white
dwarfs retain thick H layers and remain DA stars throughout their
evolution. Hence a significant fraction of the DA white dwarfs in
Figure \ref{mass_teff} should remain DA stars in the temperature range
displayed here, and eventually evolve into H-atmosphere DC stars at
low temperatures when H$\alpha$ disappears. We can see that the
coolest DA stars indeed merge with the cool DC sequence near
$\Te\sim5000$~K, giving support to the idea that most cool DC stars in
our sample have H-dominated atmospheres. Note also that in the
temperature range considered in our analysis, it is possible for some
of the DA stars to turn into He-rich DA or DC white dwarfs as a result
of convective mixing. We come back to this point below.

What is particularly noteworthy in Figure \ref{mass_teff} is the
evolution of the crystallization sequence as a function of mass and
$\Te$. The mass distribution of DA white dwarfs displayed in Figure 13
of \citet{kilic20}, based on spectroscopically confirmed DA stars
hotter than 6000 K in the 100 pc sample and the SDSS footprint, shows
a very strong peak at $\sim$0.6 \msun, and a smaller but very broad
shoulder centred at $\sim$0.8 \msun\ (see also Section
\ref{sec:evolhe} below). However, the results displayed in Figure
\ref{mass_teff} indicate that at $\Te\sim9000$~K, say, nearly half of
the DA stars lie within the crystallization boundaries near 1.0 \msun.
As we go down in temperature, crystallization occurs at lower masses,
until the crystallization sequence merges with the normal mass DA
white dwarfs around 5500 K or so. Consequently, when the cumulative mass
distribution is considered ($N$ versus $M$), the normal mass white
dwarfs accumulate in the mean peak, while the crystallized DA stars
are smeared out over this broad shoulder.

With the exception of a few objects, most white dwarfs above 0.9
\msun\ are DA stars, down to $\Te\sim5500$~K, below which they seem to
evolve into He-rich IR-faint white dwarfs. The obvious mechanism
responsible for this transformation is convective mixing.  However, as
discussed at length in \citet{bergeron22}, the small H/He abundance
ratios measured in these IR-faint white dwarfs ($\logh=-4$ to $-3$)
imply that the total amount of hydrogen in these stars is extremely
small, such that mixing should have occurred at higher temperatures
($\Te\sim9000$~K). Consequently, their immediate progenitors should be
massive DC white dwarfs instead of DA stars. Two possible explanations are
that the H abundance measured in these IR-faint white dwarfs is erroneous, or
our understanding of convective mixing needs to be revised.

\subsection{Magnetic White Dwarfs}\label{sec:evolmag}

Before discussing the evolution of He-atmosphere white dwarfs, it is
worth investigating the magnetic objects in our sample and the
role played by magnetism in the spectral evolution of these stars.
Around 5\% of the white dwarfs in our sample show signs of magnetism,
whether it is through Zeeman splitting or polarization
measurements. It is also detected in all spectral types (DA, DZ, DQ,
DC); in the case of DQ white dwarfs, the strong C$_2$ Swan bands appear
to be strongly shifted towards shorter wavelength. Various scenarios have
been invoked to explain the origin of magnetic fields in white dwarfs
(see the review by \citealt{ferrario15}), including a fossil origin,
whether it was present in the forming stellar cloud or
generated by a dynamo effect in rotating cores in the main sequence
progenitor \citep{angel81,tout04,wickramasinghe05,ferrario15}. 
There is also strong evidence that binary evolution is responsible for 
producing magnetic fields in some white dwarfs (see, e.g., \citealt{tout08,garcia12}). 
More recently, is has also been proposed that the onset of core crystallization in 
cooling white dwarfs might produce magnetic fields through a dynamo
process triggered by the onset of convection in the liquid core due to phase
separation \citep{isern17,schreiber21,ginzburg22}. It thus appears that several
mechanisms for producing magnetic fields might be at play here, which
probably vary as a function of mass and cooling age (see Figure 2 of
\citealt{bagnulo22}), or equivalently, as a function of decreasing
effective temperature.

\begin{figure*}
\centering
\includegraphics[width=4.2in, angle=270, clip=true, trim=0.in 0.in 0.0in 0.0in]{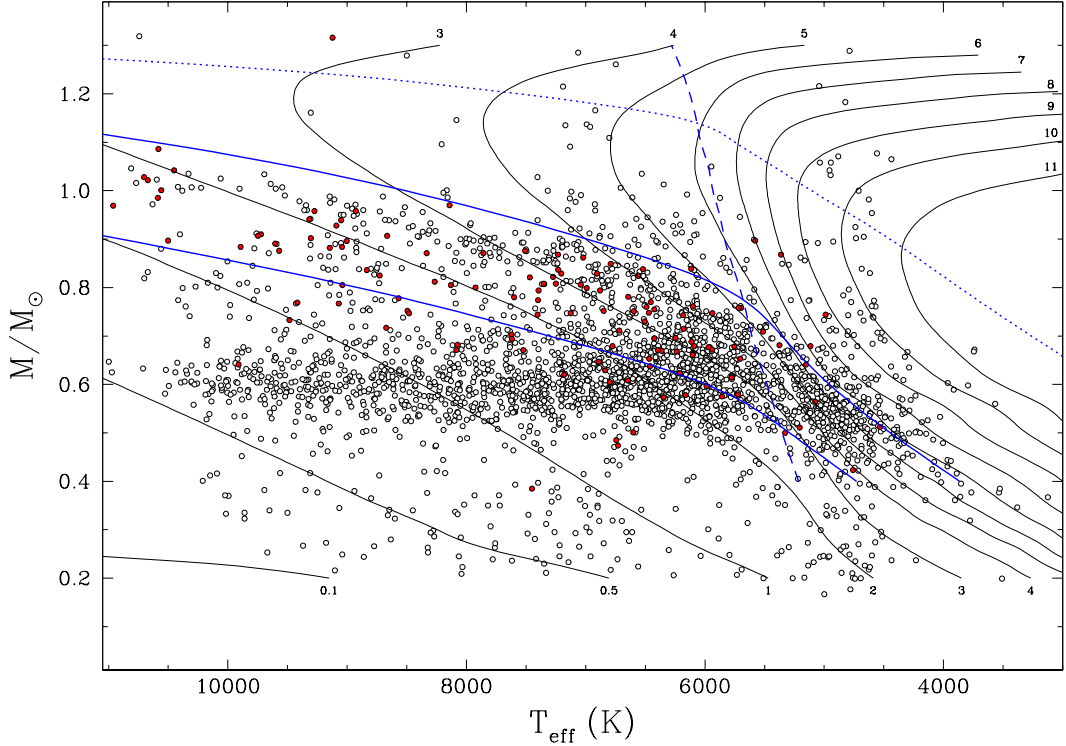}
\caption{Same as Figure \ref{mass_teff} but with magnetic white dwarfs
  shown as red dots (non-magnetic white dwarfs are shown with lighter
  circles to emphasize the magnetic objects).
\label{mass_teff_DHP}}
\end{figure*}

We show in Figure \ref{mass_teff_DHP} the same $M$ versus $\Te$
diagram as in Figure \ref{mass_teff} but this time by highlighting
with red symbols the magnetic white dwarfs in our sample. Our results
conclusively demonstrate that the presence of a magnetic field in
most, if not all white dwarfs in our sample is strongly correlated
with the phenomenon of core crystallization, as showed by the two blue
solid curves, where the lower curve indicates the onset of
crystallization at the centre of evolving models, while the upper one
indicates the locations where 80\% of the total mass has
solidified. Note that the number of magnetic white dwarfs at low
temperatures is most likely underestimated given the difficulty of
detecting Zeeman splitting in weak H$\alpha$ absorption features. The
situation is even worse in the case of DC stars, which would require
spectropolarimetric measurements \citep[e.g.,][]{putney97,bagnulo21,berdyugin22}.

It has long been suggested that convective energy transport in cool
white dwarfs ($\Te\lesssim10,000$~K) could be seriously impeded by the
presence of a strong magnetic field. \citet{tremblay15} reexamined
this problem using radiation magnetohydrodynamic simulations and
indeed showed that magnetic fields of only tens of kG could
significantly reduce the importance of convection to the point that
the atmospheres become radiative. Model atmosphere analysis of magnetic
white dwarfs can in principle be used to validate this finding by testing
whether consistent fits can be obtained with non-convective atmospheres,
but conflicting results have been reported \citep{lecavalier17,gentile18}.
In any case, magnetic fields can plausibly inhibit (at least partially)
convective transport and this may be the key physical mechanism responsible for
turning He-atmosphere white dwarfs into H-atmosphere white dwarfs at
low temperatures, a question we investigate in the next subsection.

\subsection{The Evolution of He-atmosphere White Dwarfs}\label{sec:evolhe}

\begin{figure*}
\centering
\includegraphics[width=4.2in, angle=270, clip=true, trim=0.in 0.in 0.0in 0.0in]{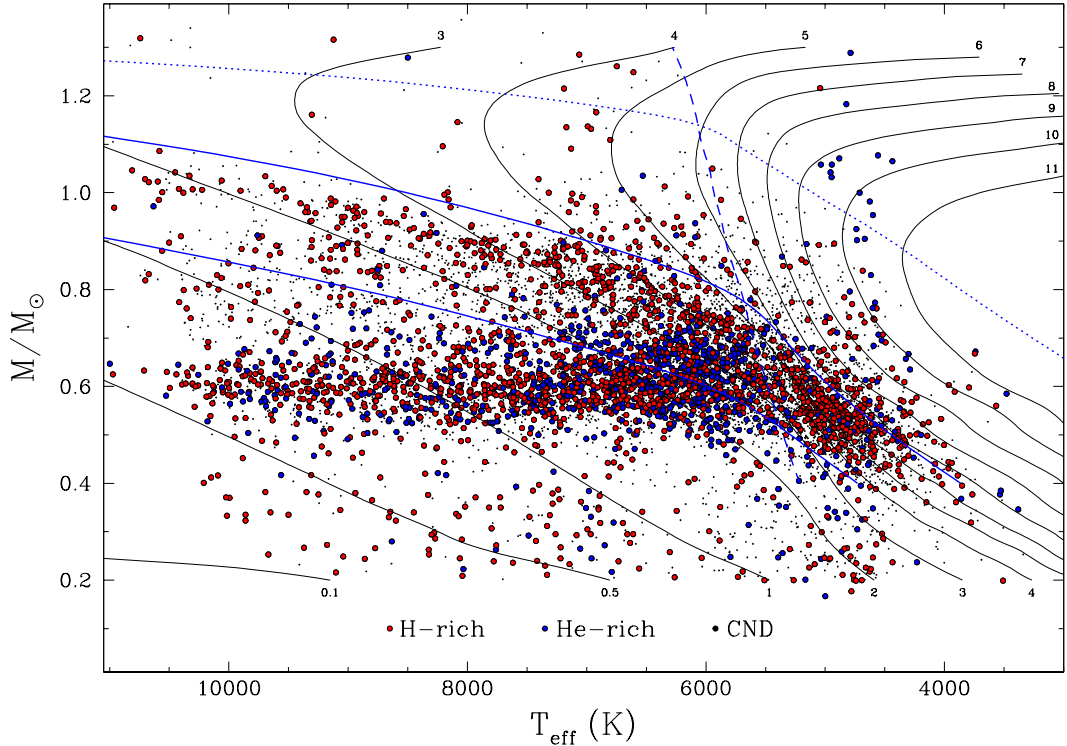}
\caption{Same as Figure \ref{mass_teff} with the exception that the
  objects are divided into H-atmosphere and He-atmosphere white
  dwarfs, as indicated in the legend.  Also shown as small dots
  are the 5358 white dwarf candidates (without spectral
  classification) analysed under the assumption of a pure H
  composition.
\label{mass_teff_atm_CND}}
\end{figure*}

We first reexamine the $M$ versus $\Te$ diagram displayed in Figure
\ref{mass_teff}, but this time, instead of differentiating the white
dwarfs in terms of their spectral type, we separate them in terms of
their main atmospheric constituent, H or He. The results are displayed
in Figure \ref{mass_teff_atm_CND}. Also indicated by small dots in
this figure are the 5358 white dwarf candidates (i.e.~without spectral
classification) found in the MWDD, and analysed under the assumption
of a pure H composition for all objects. We remind the reader that
there are only 2880 spectroscopically confirmed white dwarfs in our
100 pc sample, hence less than 35\% of the total sample.

He-atmosphere white dwarfs in our sample may have different origins.
First, their immediate progenitors could be DB stars. In this case, it
has been shown that as they cool off, the bottom of the He convection
zone can reach deep into the interior where resides a massive hydrogen
reservoir, resulting into traces of H being brought up to the surface,
and thus producing a DBA star (\citealt{rolland20}, B\'edard et
al. 2022, submitted to ApJ). In some cases, however, there may be not
enough hydrogen in the deep interior that the DB white dwarf retains a
H-free atmosphere throughout its evolution. This particular channel is
believed to be responsible for the origin of most DQ white dwarfs,
while DQ stars with traces of hydrogen, such as those analysed
in our study, are probably the result of convectively mixed DA stars
below $\Te\sim12,000$~K \citep{bedard22b}. Convective mixing has also
been invoked to account for the existence of the He-rich DA stars in
our sample, as well as those reported in \citet{rolland18}. Similarly,
He-atmosphere DZ and DZA white dwarfs may descend from DB stars, or
from convectively mixed DA stars. Finally, as discussed in Section
\ref{sec:DCcomp}, featureless DC white dwarfs must contain some traces of
hydrogen, otherwise their masses inferred from photometry are too
large; if they are devoid of hydrogen because of their
prior evolution, it is believed they will become DQ white dwarfs
instead. Thus, with the exception of some DQ stars, we expect
the vast majoriy of cool, He-atmosphere white dwarfs to contain
residual amounts of hydrogen in their atmosphere and stellar envelope. This is
obviously the case for IR-faint white dwarfs as well.

Going back to the results displayed in Figure \ref{mass_teff_atm_CND},
we must remember that a pure H composition was assumed for all DC
stars below $\Te=5200$~K. It is thus not a firm determination, but by
far the most reasonable assumption given the results of our
analysis. Indeed, as discussed in Section \ref{sec:CMD}, such a sudden
decrease of He-atmosphere white dwarfs in this temperature range is
entirely consistent with the abrupt termination of the more luminous
sequence of DC stars observed in the colour-magnitude diagram
displayed in Figure \ref{color_mag_pan}. Given the location of this
transition exactly in the range of mass and temperature where
crystallization occurs, and given the correlation between the
occurrence of magnetism and crystallization discussed in Section
\ref{sec:evolmag}, it is tantalizing to suggest that the mechanism
responsible for transforming most, but not all, He-atmosphere white
dwarfs into H-atmosphere white dwarfs is the following sequence of
events: First, crystallization occurs in the stellar core and the
crystallization front moves outward as the white dwarf cools off. This
process eventually creates a magnetic field through the dynamo
mechanism described in \citet{isern17}. Then,
the presence of the magnetic field suppresses completely, or even
partially, convective transport, allowing the hydrogen
thoroughly diluted within the mixed H/He convective envelope to
diffuse upward, gradually building a thick H layer sitting on top of a
deeper He-rich envelope. While detailed evolutionary calculations
are required to explore this suggestion more quantitatively, this
is certainly the most promising explanation for the transition from
He-atmosphere to H-atmosphere white dwarfs that occurs below
$\Te=5200$~K in Figure \ref{mass_teff_atm_CND}.  Within this scenario,
the few remaining cool ($\Te\lesssim5000$~K) He-atmosphere white
dwarfs identified in our survey, most of which are DZ stars, could be
interpreted as objects with not enough hydrogen, or no hydrogen at
all.

\begin{figure}
\centering
\includegraphics[width=3.3in, clip=true, trim=0.in 0.9in 0.0in 0.9in]{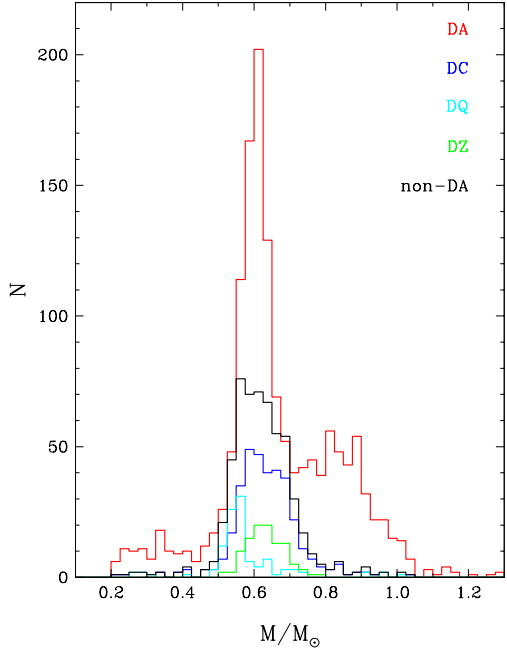}
\caption{Mass distributions of all white dwarfs above $\Te=6000$~K split into
different spectral types, as indicated in the legend. The black histogram
represents the sum of all non-DA types (DC, DQ, DZ).
\label{histo_mass}}
\end{figure}

We now consider the peak of the mass distribution at $\sim$0.6
\msun\ in Figure \ref{mass_teff_atm_CND}, but above $\Te=6000$~K where
we can easily identify DA stars. In Figure \ref{mass_teff}, we already
noticed that both the DQ and DZ white dwarfs had very narrow mass
distributions, but that the DQ sequence had a lower mean mass than the
DZ counterpart, while the DC stars appear to share the same mass
distribution as DZ white dwarfs. But if we now compare the mass
distributions of H-atmosphere and He-atmosphere white dwarfs in the
same temperature range, displayed in Figure \ref{mass_teff_atm_CND},
we notice that they are remarkably similar. We can explore this more
quantitatively by looking at the cumulative mass distributions, $N$
versus $M$, displayed in Figure \ref{histo_mass}, where we compare the
mass distribution for the H-atmosphere DA stars above 6000 K with that
for the He-atmosphere non-DA stars. For the latter, we also show the
individual contributions from DC, DQ, and DZ white dwarfs. With the
exception of the very broad shoulder observed for DA stars centred
near $\sim$0.8 \msun\ and already discussed in Section
\ref{sec:evolda}, we notice that the mass distributions of H- and
He-atmosphere white dwarfs are indeed remarkably similar, although
there are obviously more DA stars in the central peak at 0.6 \msun,
probably originating from the evolutionary channel consisting of white
dwarfs retaining thick H layers throughout their cooling history
($\sim$75\% of the entire white dwarf population according to
\citealt{bedard20}).

Given the results shown in Figure \ref{histo_mass}, is it possible
that the non-DA stars form a more homogeneous population than
previously believed? First, as discussed in \citet{bedard22b}, carbon
can be more easily detected in lower mass DQ white dwarfs since the
deep carbon diffusion tail sinks less rapidly in objects with lower
surface gravity. Second, \citet{blouin22b} demonstrated that the
simultaneous presence of metals at the photosphere of DQ
white dwarfs can change the atmospheric structure significantly, to
the point that carbon features vanish, thus explaining the paucity
of DQZ white dwarfs. This could partially explain
the near absence of an overlap between the mass distributions of DQ and DZ
white dwarfs, as observed in both Figures \ref{mass_teff} and
\ref{histo_mass}. Finally, given that DC and DZ white dwarfs probably
differ only by the presence of surrounding material being accreted,
it is then possible that a significant fraction of DC
stars also result from the evolution of this so-called PG
1159--DO--DB--DQ scenario (see \citealt{bedard22b} and references
therein), and that only a small fraction of non-DA stars above 6000 K
originate from convectively mixed DA white dwarfs.

\begin{figure}
\centering
\includegraphics[width=3.3in, clip=true, trim=0.in 0.5in 0.0in 0.5in]{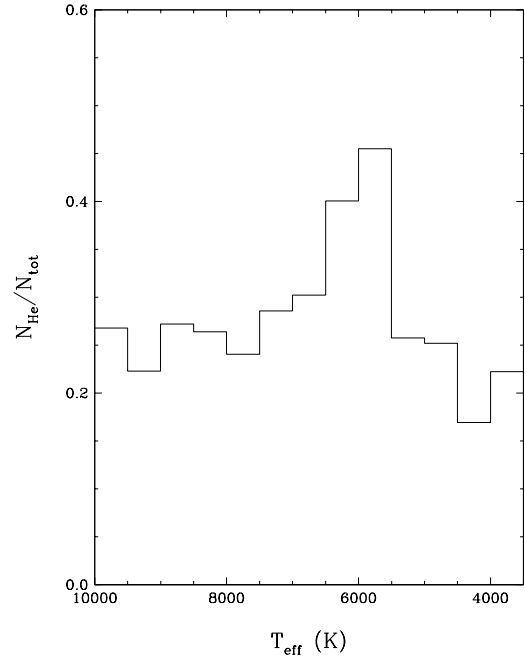}
\caption{Ratio of He-atmosphere white dwarfs to the total number of stars
  as a function of effective temperature.
\label{histo_fraction}}
\end{figure}

Although our sample is not statistically complete in any sense, we can
still test this last hypothesis by looking at the variation of the
fraction of He-atmosphere white dwarfs as a function of $\Te$, the
results of which are displayed in Figure \ref{histo_fraction}.
Surprisingly, despite the ill-defined statistical properties of our
sample, our results are entirely consistent with those displayed in
Figure 8 of \citet{mccleery20}, which are based on the volume-limited
40 pc sample in the northern hemisphere. First, we can see that above
$\Te\sim6500$~K, the fraction of He-atmosphere white dwarfs remains
nearly constant around 25\%, which suggests that there is little
evidence for convective mixing of DA white dwarfs into non-DA stars,
at least not in the range of $\Te$ considered in our analysis.  The
situation changes drastically at lower temperatures, however, where
the fraction increases to $\sim$45\% around 6500 K, and gradually
decreases to 15\% around 4000 K. We have to be careful, however, not
to overinterpret our results at very low $\Te$ values because this is
the range of temperature where our spectro-photometric analysis is
most uncertain, and also because this is where the fraction of white
dwarf candidates (i.e., not spectroscopically confirmed) is the
largest. Indeed, from the results displayed in Figure
\ref{mass_teff_atm_CND}, we estimate that above 6000 K, 40\% of our
total sample of objects are spectroscopically confirmed white dwarfs,
while this fraction drops to $\sim$25\% below this temperature.

We thus conclude from the above discussion that convective mixing
occurs in DA white dwarfs, but mostly at low effective temperature. In
turn, this implies that most DA stars in the temperature range
considered in our analysis have thick hydrogen layers around
$\log\qh\sim-7.5$ (at $M=0.6$ \msun) according to the results
displayed in Figure 19 of \citet{bergeron22}, which explores the
expected variation of the H/He ratio as a function of $\Te$ for DA
models with various hydrogen layer masses.

\section{Conclusions}\label{sec:conclusion}

We selected 8238 white dwarfs and white dwarf candidates from the MWDD
within 100 pc from the Sun and below $\Te\sim10,000$K, 2880 of which
were spectroscopically confirmed degenerates (35\% of the
sample). Optical and infrared photometry as well as spectroscopic
observations, when available, were combined to measure the stellar
parameters for each individual white dwarf in our sample using
state-of-the-art model atmospheres appropriate for each spectral
type. We can draw the following conclusions from our analysis:

1. There is now strong evidence that most, but probably not all, cool
DC white dwarfs have H atmospheres. This conclusion is supported not
only by the location of these objects in colour-magnitude diagrams,
but also from our detailed photometric analysis. Effective temperature
and stellar masses inferred from He-atmosphere models are
unrealistically too low. However, our analysis has also revealed
problems with the pure H model atmospheres, both in color-magnitude
diagrams and photometric fits, which we attributed to inaccuracies in
the calculations of opacity sources, whether it is the red wing of
L$\alpha$, the H$^-$ bound-free opacity, or the CIA opacity from
molecular hydrogen.

2. There is a more luminous observed sequence of DC white dwarfs in
colour-magnitude diagrams that terminates around $\Te\sim5200$~K. Our
detailed photometric analysis suggests that below this temperature,
He-atmosphere white dwarfs turn into H-atmosphere white dwarfs. The
mechanism by which this transformation occurs appears to be related to
the onset of crystallization and to the occurrence of magnetism being
generated as a result of this process. The presence of a magnetic
field would lead to a suppression, even partially, of convective
transport, allowing hydrogen, thoroughly diluted within the deep
stellar envelope, to float back to the surface.

3. Some DQ white dwarfs show traces of hydrogen in the form of an
H$\alpha$ absorption feature. In some cases, these are obviously
unresolved DA + DQ binaries (due in part to their overluminosity), but
other objects are genuine DQA stars, most likely originating from
convectively mixed DA stars. We believe that we may also have
identified a population of cool DQ white dwarfs in which the presence of
hydrogen manifests itself as collision-induced absorption by molecular
hydrogen in the near-infrared, although a more detailed analysis using
model atmospheres including H, He, and C is required to confirm this
interpretation.

4. We identified several unresolved double degenerate binaries in
our sample, all of which have been analyzed under the assumption
of a single object. These systems deserve further study to deconvolve
the stellar parameters of the individual components, following
the approach described in \citet{bedard17}.

5. Our analysis reveals that most low-mass white dwarfs probably
have H atmospheres, which would in turn indicate that common-envelope
evolution most likely produces white dwarf remnants that retain thick
H layers.

6. The mass distributions of DQ, DZ, and DC white dwarfs suggest that
these may form a more homogeneous population than previously believed,
with the DQ stars representing the low-mass component of the non-DA
white dwarfs, in which carbon features can be more easily
detected. Carbon features would be more difficult to detect in DZ and
DC white dwarfs due to their higher mass and/or larger metal
content. If this is indeed the case, it is possible that a larger
fraction of non-DA white dwarfs below $\Te=10,000$~K evolve from the
so-called PG 1159--DO--DB--DQ scenario.

7. We found little evidence in our sample for the transformation of a
large fraction of DA stars into He-atmosphere white dwarfs through the
process of convective mixing between $\Te=10,000$~K and $\sim$6500
K. The He-rich DA stars are the obvious exception, but there are very
few of those in our sample (or even in the entire SDSS white dwarf
sample for that matter, see \citealt{rolland18}). 

8. We do, however, find evidence for convective mixing below
$\Te\sim6500$~K, where the fraction of He-atmosphere white dwarfs
increases to $\sim$45\%, which suggests that most DA stars in the
temperature range considered in our analysis have thick hydrogen
layers around $\log\qh\sim-7.5$.

9. Finally, there is an obvious need for more near-infrared photometry, with
better accuracy and precision, required for the analysis of cool
($\Te<6000$~K) white dwarfs, in particular for the determination of
the chemical composition, and in some cases even the main atmospheric
constituent.

Our understanding of the spectral evolution of cool white dwarfs is like a
giant puzzle, where individual pieces are added one by one in order to
build a better picture, as better observational data and theoretical
models become available over the years. We feel we have added a few
pieces of that puzzle in our analysis, but the overall picture remains
incomplete. Even though we are convinced that progess can only be
accomplished through tailored analyses of individual objects in a
given sample, we felt we have reached the limit of human capacity to
analyse individually each object in such a large sample of nearly 3000
white dwarfs. Better techniques for handling bigger data sets
involving machine learning algorithms will eventually become necessary.

\section*{Acknowledgements}

We are grateful to A. B\'edard for enlightening discussions. This work
was supported in part by NSERC Canada and by the Fund FRQ-NT
(Qu\'ebec). SB is a Banting Postdoctoral Fellow and a CITA National
Fellow, supported by NSERC.

Funding for the Sloan Digital Sky Survey (\url{https://www.sdss.org})
has been provided by the Alfred P. Sloan Foundation, the
U.S. Department of Energy Office of Science, and the Participating
Institutions. SDSS-IV acknowledges support and resources from the
Center for High-Performance Computing at the University of Utah, and
is managed by the Astrophysical Research Consortium for the
Participating Institutions of the SDSS Collaboration
\citep{blanton17}.

This work has made use of data from the European Space Agency mission
{\it Gaia} (\url{https://www.cosmos.esa.int/gaia}), processed by the
{\it Gaia} Data Processing and Analysis Consortium (DPAC). Funding for
the DPAC has been provided by national institutions, in particular the
institutions participating in the {\it Gaia} Multilateral Agreement
\citep{gaia18b}.

This research has also made use of the NASA Astrophysics Data System
Bibliographic Services; the Montreal White Dwarf Database
\citep{dufour17}; the SIMBAD database, operated at the Centre de
Donn\'ees astronomiques de Strasbourg \citep{wenger00}; and the
NASA/IPAC Infrared Science Archive, operated at the California
Institute of Technology.

\section*{Data Availability}

The astrometric, photometric and spectroscopic data underlying this
article are available in the MWDD at
http://www.montrealwhitedwarfdatabase.org.  The astrometric and
photometric data are also available as online supplementary material.

\bibliography{ms}{}
\bibliographystyle{mnras}

\bsp
\label{lastpage}

\end{document}